\documentclass[11pt,english]{article}
\usepackage[T1]{fontenc}
\usepackage[latin9]{inputenc}
\usepackage{geometry}
\usepackage{color}
\usepackage{array}
\usepackage{float}
\usepackage{mathtools}
\usepackage{multirow}
\usepackage{amsmath}
\usepackage{amssymb}
\usepackage{graphicx}
\usepackage{setspace}
\usepackage{wasysym}
\usepackage[authoryear]{natbib}
\makeatletter

\providecommand{\tabularnewline}{\\}
\floatstyle{ruled}
\newfloat{algorithm}{tbp}{loa}
\providecommand{\algorithmname}{Algorithm}
\floatname{algorithm}{\protect\algorithmname}

\@ifundefined{date}{}{\date{}}
\usepackage{enumitem}
\setlist{nolistsep}

\bibpunct{(}{)}{,}{autheryear}{,}{,}

\newtheorem{assumption}{Assumption}
\newtheorem{theorem}{Theorem}

\newtheorem{remark}{Remark}
\newtheorem{example}{Example}

\newtheorem{proposition}{Proposition}

\newcommand*{\indep}{%
  \mathbin{%
    \mathpalette{\@indep}{}%
  }%
}
\newcommand*{\nindep}{%
  \mathbin{
    \mathpalette{\@indep}{\not}
  }%
}
\newcommand*{\@indep}[2]{%
  \sbox0{$#1\perp\m@th$}
  \sbox2{$#1=$}
  \sbox4{$#1\vcenter{}$}
  \rlap{\copy0}
  \dimen@=\dimexpr\ht2-\ht4-.2pt\relax
  \kern\dimen@
  {#2}%
  \kern\dimen@
  \copy0 
} 

\newcommand{\pr}{{P}}
\newcommand{\bP}{\mathbb{P}}
\newcommand{\G}{\mathbb{G}}
\newcommand{\E}{\mathbb{E}}
\newcommand{\var}{\mathrm{var}}

\newcommand{\R}{\mathbb{R}}
\newcommand{\de}{\mathrm{d}}
\newcommand{\T}{\mathrm{\scriptscriptstyle T}}
\newcommand{\logit}{\text{logit}}

\newcommand{\temp}{\mathrm{temp}}

\newcommand{\cont}{\mathrm{cont}}
\newcommand{\disc}{\mathrm{disc}}

\newcommand{\dur}{\mathrm{dur}}

\newcommand{\eff}{\mathrm{eff}}

\usepackage{babel}

\makeatother

\usepackage{babel}
\begin{document}

\global\long\def\spacingset#1{\global\long
\global\long\def\baselinestretch{%
}%
\small\normalsize}%
 \spacingset{1}

\title{\textbf{Structural nested mean models with irregularly spaced longitudinal
observations}}
\author{Shu Yang\thanks{Department of Statistics, North Carolina State University, North Carolina
27695, U.S.A. Email: syang24@ncsu.edu}}

\maketitle
\bigskip{}

\begin{abstract}
Structural Nested Mean Models (SNMMs) are useful for causal inference
of treatment effects in longitudinal observational studies. Most existing
works assume that the data are collected at pre-fixed time points
for all subjects, which, however, is restrictive in practice. To deal
with irregularly spaced observations, we assume a class of continuous-time
SNMMs and a martingale condition of no unmeasured confounding (NUC)
to identify the causal parameters. We develop the first semiparametric
efficiency theory and locally efficient estimators for continuous-time
SNMMs. This task is non-trivial due to the restrictions from the NUC
assumption imposed on the SNMM parameter. In the presence of dependent
censoring, we propose an inverse probability of censoring weighting
estimator, which achieves a multiple robustness feature in that it
is unbiased if either the model for the treatment process or the potential
outcome mean function is correctly specified, regardless whether the
censoring model is correctly specified. The new framework allows us
to conduct causal analysis respecting the underlying continuous-time
nature of the data processes. We estimate the effect of time to initiate
highly active antiretroviral therapy on the CD4 count at year 2 from
the observational Acute Infection and Early Disease Research Program
database.
\end{abstract}

\noindent \vfill{}

\newpage\spacingset{1.45} 

\section{Introduction\label{sec:Introduction}}

\subsection{Causal inference methods with time-varying confounding}

The gold standard to draw causal inference of treatment effects is
designing randomized experiments. However, randomized experiments
are not always feasible due to practical constraints or ethical issues.
Moreover, randomized experiments often have restrictive inclusion
and exclusion criteria for patient enrollment, which limits the experiment
results to be generalized to a larger real-world patient population.
In these cases, observational studies are useful. In observational
studies, confounding by indication poses a unique challenge to drawing
valid causal inference of treatment effects. For example, sicker patients
are more likely to take the active treatment, whereas healthier patients
are more likely to take the control treatment. Consequently, it is
not fair to compare the outcome from the treated group and the control
group directly. Moreover, in longitudinal observational studies, confounding
is likely to be time-dependent, in the sense that time-varying prognostic
factors of the outcome affect the treatment assignment at each time,
and thereby distort the association between treatment and outcome
over time. In these cases, the traditional regression methods are
biased even adjusting for the time-varying confounders \citep{robins1992g,hernan2000marginal,hernan2005structural,robins2009estimation,orellana2010dynamic_a}.

\subsection{A motivating application}

\textbf{}HAART (highly active antiretroviral therapy) is the standard
of care as initial treatment for HIV. Our interest is motivated by
the observational AIEDRP (Acute Infection and Early Disease Research
Program) Core 01 study. This study established a cohort of HIV infected
patients who have chosen to defer therapy but agree to be followed
by this study. Deferring therapy may have an increased risk of permanent
immune system damage but also a decreased risk of developing drug
resistance. We aim to determine the effect of time to initiate HAART
on disease progression for those patients who were diagnosed during
acute or early HIV infection.

The outcome variable $Y$ is the CD4 count measured by the end of
year $2$, for which lower counts indicate worse immunological function
and disease progression. The inter-quantile range of the observed
outcome in the AIEDRP database is from $443$ cells/mm$^{3}$ to $794$
cells/mm$^{3}$. In this database, $45\%$ of patients dropped out
of the study before year 2, rendering $969$ patients with complete
observations. Treatment initiation can only occur at follow-up visits
and be determined by the discretion of physicians. By protocol, follow-up
visits occur at weeks 2, 4, and 12, and then every 12 weeks thereafter,
through week 96. However, as shown in Figure \ref{fig:irregular visit},
both the number and the timings of visits differ from one patient
to the next. Among all patients, $36\%$ of patients did not initiate
the treatment before year 2. The observed time to treatment initiation
ranges continuously from $12$ days to $282$ days. The covariates
include age at infection, gender, race, injection drug ever/never,
and measured CD4 count and log viral load at follow-up visits.

To answer the question of interest using the AIEDRP database, two
major concerns arise: first, the association between the treatment
and outcome processes, i.e., time-varying confounding, that would
obscure the causal effect of time to treatment initiation on the CD4
outcome at year 2; second, the observations are irregularly spaced.

\subsection{Structural nested mean models}

Structural Nested Models (SNMs; \citealp{robins1992g,robins1994correcting})
have been proposed to overcome the challenges for causal inference
with time-varying confounding. We focus on a class of SNMs for continuous
outcomes, namely, structural nested mean models (SNMMs). We discuss
the extension to accommodate the binary outcome and the survival outcome
in Section \ref{sec:Discussion}. Most existing works on SNMMs assume
discrete-time data generating processes and require all subjects to
be followed at the same pre-fixed time points, such as months. The
literature of discrete-time SNMMs is fruitful; see, e.g., \citet{robins1998structural,robins2000sensitivity,almirall2010structural,chakraborty2013statistical,lok2012impact,lok2014opt,yang2015gof,yang2017sensitivity}.
However, as in the AIEDRP database, observational data are often collected
by user-initiated visits to clinics, hospitals and pharmacies, and
data are more likely to be measured at irregularly spaced time points,
which are not necessarily the same for all subjects. \textcolor{black}{Such
data sources are now commonplace}, such as electronic health records,
claims databases, disease data registries, and so on \citep{chatterjee2016constrained}.

The existing causal framework does not directly apply in such situations,
requiring some (possibly arbitrary) discretization of the timeline
\citep{neugebauer2010observational}. Such data pre-processing is
quite standard and routine to practitioners, but leads to many unresolved
problems: the treatment process depends transparently on the discretization,
and therefore the interpretation of SNMMs depends on the definition
of time interval \citep{robins1998correction}. \textcolor{black}{Moreover,
after discretization, the data may need to be recreated at certain
time points. Consider monthly data for example. If a subject had multiple
visits within the same month, a common strategy is to take the average
of the multiple measures as the observation for a given variable at
that month. If a subject had no visit for a given month, one may need
to impute the missing observation. Because of such distortions, the
resulting data may not satisfy the standard causal consistency or
no unmeasured confounding (NUC) assumptions. Consequently, model parameters
may not have a causal interpretation.}

With irregularly spaced observations, it is more reasonable to assume
that the data are generated from continuous-time processes. The work
for causal models in continuous-time processes is somewhat sparse;
exceptions include, e.g., \citet{robins1998correction,lok2004estimating,lok2008statistical,zhang2011causal,lok2017mimicking}.
Extending the existing causal models with discrete-time processes
to continuous-time processes is not trivial. An important challenge
lies in time-dependent selection bias or confounding; e.g., in a health-related
study, sicker patients may visit the doctor more frequently and are
more likely to initiate the treatment. To overcome this challenge,
following \citet{lok2008statistical}, we treat the observed treatment
assignment process as a counting process $N_{T}(t)$ and assume a
martingale condition of NUC on $N_{T}(t)$ to identify the SNMM parameters.
Specifically, the NUC assumption entails that the jumping rate of
$N_{T}(t)$ at $t$ does not depend on future potential outcomes,
given the past treatment and covariate history up to $t$. A practical
implication is that the covariate set should be rich enough to include
all predictors of outcome and treatment, so that we can distinguish
the treatment effect and the confounding effect. This assumption was
also adopted in \citet{zhang2011causal} and \citet{yang2018modeling}
to the settings where the effect of a treatment varies in continuous
time. \citet{lok2017mimicking} provided a strategy of constructing
unbiased estimating equations exploiting the relationship between
the mimicking potential outcome process and the treatment process,
which leads to a large class of estimators. While this strategy provides
unbiased estimators, there is no guidance on how to choose an efficient
estimator, and a naive choice can lead to inaccurate estimation.

\subsection{Semiparametric efficiency theory for continuous-time SNMMs}

We establish the new semiparametric efficiency theory for continuous-time
SNMMs with irregularly spaced observations. Toward this end, we follow
the geometric approach of \citet{bickel1993efficient} for the semiparametric
model by characterizing the nuisance tangent space, its orthogonal
complementary space, and lastly the semiparametric efficiency score
for the SNMM parameter.

In our problem, the SNMM and the NUC assumption constitute the semiparametric
model for the data. Given the close relationship of causal inference
and missing data theory, it is worthwhile to discuss the connection
of the semiparametric efficiency development in our paper and that
in the missing data literature \citep{ding2017causal}. The NUC assumption
for the treatment process plays the same role of the ignorability
assumption for the missing data mechanism; therefore, our characterization
of the nuisance tangent space for the treatment process follows the
same as that for the continuous-time missing data process; see Section
5.2 of \citet{tsiatis2007semiparametric}. Besides this analogy, our
theoretical task is \textcolor{black}{considerably more complicated}.
Although the NUC assumption does not have any testable implications
on the observed-data likelihood \citep{van2003unified,tan2006regression},
it imposes conditional independence restrictions on the treatment
process and the counterfactual outcomes, given the past history, and
hence restrictions for the SNMM parameter; see equation (\ref{eq:UNC2}).
To circumvent this complication, we use the variable transformation
technique and translate the restrictions into the new variables, which
leads to the unconstrained observed data-likelihood. This step allows
us to characterize the semiparametric efficiency score for the SNMM
parameter and construct locally efficient estimators which achieve
the semiparametric efficiency bound.

In the AIEDRP database, a large portion of patients dropped out of
the study before year 2. To accommodate possible dependent censoring
due to drop-out, we propose the inverse probability of censoring weighting
(IPCW) estimator. We show that the proposed estimator is multiply
robust in that it is consistent if either the potential outcome mean
model is correctly specified or the model for the treatment process
is correctly specified, regardless whether the censoring model is
correctly specified. This amounts to six scenarios specified in Table
\ref{tab:Multiply-Robustness} that guarantee consistent estimation,
allowing some components in the union of the three models to be misspecified
\citep{molina2017multiple,wang2018bounded}. Moreover, using the empirical
process theory \citep{van1996weak}, we characterize the asymptotic
property of the proposed estimator of the SNMM parameter under a parametric
outcome mean model, and proportional hazards models for the treatment
and censoring processes, allowing for multiply robust inference.

It is important to note that for regularly spaced observations, i.e.
the data process can only take values at pre-fixed time points, the
proposed estimator simplifies to the existing estimator with discrete-time
data. For irregularly spaced observations, the new model and estimation
framework allows us to deal with irregularly spaced observations directly
and respects the nature of the underlying data generating mechanism.
In contrast, the existing g-estimator requires data pre-processing
and may introduce bias as demonstrated by simulation in Section \ref{sec:simulation}.

The rest of the article is organized as follows. In Section \ref{sec:discrete-SNMM},
we describe the SNMM with discrete-time processes, which serves as
a building block to establishing the semiparametric efficiency theory
for continuous-time processes and also enables us to establish their
connection. In Section \ref{sec:continuous SNMM}, we present the
semiparametric efficiency theory and locally efficient estimators
for the continuous-time SNMM under the NUC assumption. Moreover, we
propose an IPCW estimator to deal with dependent censoring due to
premature dropout. In Section \ref{sec:Asymptotic-property}, we establish
the asymptotic property of the estimator allowing for multiply robust
inference. In Section \ref{sec:simulation}, we present simulation
studies to investigate the performance of the proposed estimator compared
to the existing competitor in finite samples. In Section \ref{sec:Application},
we apply the proposed estimator to estimate the effect of the time
between HIV infection and initiation of HAART on the CD4 count at
year 2 after infection in HIV-positive patients with early and acute
infection. We conclude the article with discussions in Section \ref{sec:Discussion}.

\section{Structural nested mean models in discrete-time processes\label{sec:discrete-SNMM}}

\subsection{Setup, models, and assumptions}

\textcolor{black}{We first describe the SNMM in discrete-time processes.
}We assume that $n$ subjects are followed at pre-fixed discrete times
$t_{0}<\cdots<t_{K+1}$ with \textcolor{black}{$t_{0}=0$ and $t_{K+1}=\tau$.}
We assume that the subjects are simple random samples from a larger
population\textcolor{black}{{} \citep{rubin1978bayesian}. For simplicity,
we suppress the subscript $i$ for subjects.} Let $L_{m}$ be a vector
of covariates at time $t_{m}$. Let $A_{m}$ be the treatment indicator
at $t_{m}$; i.e., $A_{m}=1$ if the subject was on treatment at $t_{m}$
and $A_{m}=0$ otherwise. We use the overline notation to denote a
variable's history; e.g., $\overline{A}_{m}=(A_{0},\ldots,A_{m})$.
We assume that once treatment is initiated, it is never discontinued,
so each treatment regime corresponds to one treatment initiation time.
Let $T$ be the time to treatment initiation, and let $T=\infty$
if the subject never initiated the treatment during the follow up.
Let $\Gamma$ be the indicator that the treatment initiation time
is less than $\tau$; i.e., $\Gamma=1$ if the subject initiated the
treatment before $\tau$ and $\Gamma=0$ otherwise. \textcolor{black}{Let
$Y^{(m)}$ be the potential outcome at the end of study $\tau$, had
the subject initiated the treatment at $t_{m}$, }and let $Y^{(\infty)}$
\textcolor{black}{be the potential outcome at $\tau$ had the subject
never initiated the treatment during the study follow up. }Let $V_{m}=(A_{m-1},L_{m})$
be the vector of treatment and covariate. Let $Y$ be the continuous
outcome measured at $\tau$. Finally, the subject's full record is
$F=(\overline{A}_{K},\overline{L}_{K},Y)$. 

Following \citet{lok2012impact}, we describe the discrete-time SNMM
for the treatment effect as follows.

\begin{assumption}[Discrete-time SNMM]\label{asump:disc-SNMMs}For
$0\leq m\leq K$, the discrete-time SNMM is 
\begin{equation}
\gamma_{m}(\overline{L}_{m})=\E\left\{ Y^{(m)}-Y^{(\infty)}\mid\overline{A}_{m-1}=\overline{0},\overline{L}_{m}\right\} =\gamma_{m}(\overline{L}_{m};\psi^{*});\label{eq:disc-SNMM}
\end{equation}
i.e., $\gamma_{m}(\overline{L}_{m};\psi)$ with $\psi\in$$\R^{p}$
is a correctly specified model for $\gamma_{m}(\overline{L}_{m})$
with the true parameter value $\psi^{*}.$

\end{assumption}

This model specifies the conditional expectation of the treatment
contrasts $Y^{(m)}-Y^{(\infty)}$, given subject's observed treatment
and covariates history $(\overline{A}_{m-1}=\overline{0},\overline{L}_{m})$.
Intuitively, it states that the conditional mean of the outcome is
shifted by $\gamma_{m}(\overline{L}_{m};\psi^{*})$ had the subject
initiated the treatment at $t_{m}$ comparing to never starting. Therefore,
the parameter $\psi^{*}$ has a causal interpretation. To help understand
the model, consider $\gamma_{m}(\overline{L}_{m};\psi^{*})=(\psi_{1}^{*}+\psi_{2}^{*}t_{m})(\tau-t_{m})$,
where $\psi^{*}=(\psi_{1}^{*},\psi_{2}^{*})$. This model entails
that on average, the treatment would increase the mean of the outcome
had the subject initiated the treatment at  $t_{m}$ by $(\psi_{1}^{*}+\psi_{2}^{*}t_{m})(\tau-t_{m})$,
and the magnitude of the increase depends on the duration of the treatment
and the treatment initiation time. If $\psi_{1}^{*}+\psi_{2}^{*}t_{m}>0$
and $\psi_{2}^{*}<0$, it indicates the treatment is beneficial and
earlier initiation is better.

We make the consistency assumption to link the observed data to the
potential outcomes.

\begin{assumption}[Consistency]\label{asump:(Consistency)}The observed
outcome is equal to the potential outcome under the actual treatment
received; i.e., $Y=Y^{(T)}$.

\end{assumption}

If all potential outcomes were observed for each subject, we can directly
compare these outcomes to infer the treatment effect; however, the
fundamental problem in causal inference is that we can not observe
all potential outcomes for a particular subject \citep{holland1986statistics}.
In particular, we can observe $Y^{(\infty)}$ only for the subjects
who did not initiate the treatment during the follow up. To overcome
this issue, we define 
\begin{equation}
H(\psi^{*})=Y-\gamma_{T}(\overline{L}_{T};\psi^{*}).\label{eq:def of H}
\end{equation}
Intuitively, $H(\psi^{*})$ subtracts the treatment effect $\gamma_{T}(\overline{L}_{T};\psi^{*})$
from the observed outcome $Y$, so it mimics the potential outcome
$Y^{(\infty)}$ had the treatment never been initiated. We provide
the formal statement as proved in \citet{lok2012impact}.

\begin{proposition}[Mimicking $Y^{(\infty)}$]\label{(Mimicking-counterfactual-outcomes)}Under
Assumption \ref{asump:(Consistency)}, $H(\psi^{*})$ mimics $Y^{(\infty)}$,
in the sense that
\[
\E\left\{ H(\psi^{*})\mid\overline{A}_{m-1}=\overline{0},A_{m},\overline{L}_{m}\right\} =\E\left\{ Y^{(\infty)}\mid\overline{A}_{m-1}=\overline{0},A_{m},\overline{L}_{m}\right\} ,
\]
for $0\leq m\leq K,$ where by convention, $\E\left(\cdot\mid\overline{A}_{-1}=\overline{0},A_{0},\overline{L}_{0}\right)=\E\left(\cdot\mid A_{0},\overline{L}_{0}\right)$.

\end{proposition}

We can not fit the SNMM by a regression model pooled over time, because
the model involves the unobserved potential outcomes. Parameter identification
requires the NUC assumption \citep{robins1992g}.

\begin{assumption}[No unmeasured confounding]\label{asump:(No-unmeasured-confounding)}$A_{m}\indep Y^{(\infty)}\mid(\overline{A}_{m-1},\overline{L}_{m})$
for $0\leq m\leq K$, where $\indep$ means ``is (conditionally)
independent of'' \citep{dawid1979conditional}.

\end{assumption}

Assumption \ref{asump:(No-unmeasured-confounding)} holds if $(\overline{A}_{m-1},\overline{L}_{m})$
contains all prognostic factors for $Y^{(\infty)}$ that affect the
treatment decision at  $t_{m}$ for $0\leq m\leq K$. Under this assumption,
the observational study can be conceptualized as a sequentially randomized
experiment.

Proposition \ref{(Mimicking-counterfactual-outcomes)} implies that
under Assumption \ref{asump:(No-unmeasured-confounding)}, for $0\leq m\leq K$,
\textit{ 
\begin{equation}
\E\left\{ H(\psi^{*})\mid\overline{A}_{m-1}=\overline{0},A_{m},\overline{L}_{m}\right\} =\E\left\{ H(\psi^{*})\mid\overline{A}_{m-1}=\overline{0},\overline{L}_{m}\right\} ;\label{eq:model-part1}
\end{equation}
}see, e.g., \citet{robins1992g,lok2004estimating,lok2012impact}.
Equation (\ref{eq:model-part1}) also poses restrictions for $\psi^{*}$.

\subsection{Semiparametric efficiency theory}

The semiparametric model is characterized by the discrete-time SNMM
(\ref{eq:disc-SNMM}) and restriction (\ref{eq:model-part1}), where
the parameter of primary interest is $\psi^{*}$.

We first present the general semiparametric efficiency theory. Suppose
the data consist of $n$ independent and identically distributed random
variables $F_{1},\ldots,F_{n}$. We consider regular asymptotically
linear (RAL) estimators $\widehat{\psi}_{n}$ for $\psi^{*}$ as 
\begin{equation}
n^{1/2}(\widehat{\psi}_{n}-\psi^{*})=n^{1/2}\bP_{n}\Phi(F)+o_{p}(1),\label{eq:linear}
\end{equation}
where $\bP_{n}$ denotes the empirical mean; i.e., $\bP_{n}\Phi(F)=n^{-1}\sum_{i=1}^{n}\Phi(F_{i})$,
$\Phi(F)$ is called the influence function of $\widehat{\psi}_{n}$,
with mean zero and finite and non-singular variance. Because $\psi^{*}$
is $p$-dimensional, $\Phi(F)$ is also $p$-dimensional. From (\ref{eq:linear}),
the asymptotic variance of $n^{1/2}(\widehat{\psi}_{n}-\psi^{*})$
is equal to the variance of its influence function. As a result, to
construct the efficient RAL estimator, it suffices to find the influence
function with the smallest variance.

To do this, we take a geometric approach of \citet{bickel1993efficient}.
Consider the Hilbert space $\mathcal{H}$ of all $p$-dimensional,
mean-zero finite variance measurable functions of $F$, denoted by
$h(F)$, equipped with the covariance inner product $<h_{1},h_{2}>=\E\left\{ h_{1}(F)^{\T}h_{2}(F)\right\} $
and the norm $||h||=\E\left\{ h(F)^{\T}h(F)\right\} ^{1/2}<\infty$.
\citet{bickel1993efficient} stated that influence functions for RAL
estimators lie in the orthogonal complement of the nuisance tangent
space in $\mathcal{H}$. To motive the concept of the nuisance tangent
space for a semiparametric model, we first consider a fully parametric
model $f(F;\psi,\theta)$, where $\psi$ is a $p$-dimensional parameter
of interest, and $\theta$ is an $q$-dimensional nuisance parameter.
The score vectors of $\psi$ and $\theta$ are $S_{\psi}(F)=\partial\log f(F;\psi,\theta^{*})/\partial\psi$
and $S_{\theta}(F)=\partial\log f(F;\psi^{*},\theta)/\partial\theta$,
both evaluated at the true values $(\psi^{*},\theta^{*})$, respectively.
For a parametric model, the nuisance tangent space $\Lambda$ is the
linear space in $\mathcal{H}$ spanned by the $q$-dimensional nuisance
score vector $S_{\theta}(F)$. For semiparametric models, where the
nuisance parameter is infinite-dimensional, the nuisance tangent space
$\Lambda$ is defined as the mean squared closure of all parametric
sub-model nuisance tangent spaces. The efficient score $S_{\mathrm{eff}}(F)$
for the semiparametric model is the projection of $S_{\psi}$ onto
the orthogonal complementary space of the nuisance tangent space $\Lambda^{\bot}$;
i.e., $S_{\mathrm{eff}}(F)=\prod\left(S_{\psi}\mid\Lambda^{\bot}\right)$,
where $\prod$ is the projection operator in the Hilbert space. The
efficient influence function is $\Phi_{\mathrm{eff}}(F)=\left[\E\left\{ S_{\mathrm{eff}}(F)S_{\mathrm{eff}}(F)^{\T}\right\} \right]^{-1}S_{\mathrm{eff}}(F)$,
with the variance $\left[\E\left\{ S_{\mathrm{eff}}(F)S_{\mathrm{eff}}(F)^{\T}\right\} \right]^{-1}$,
which achieves the semiparametric efficiency bound \citep{bickel1993efficient}.
From this geometric point of view, to derive efficient semiparametric
estimators for $\psi^{*}$, it suffices to find the efficient score
$S_{\mathrm{eff}}(F)$.

\subsection{Influence functions}

The key step is to characterize the space where the influence functions
of RAL estimators belong to, i.e., the orthogonal complementary space
of the nuisance tangent space $\Lambda^{\bot}$. Following \citet{robins1994correcting},
Proposition \ref{Prop: discr-nuisance} characterizes all influence
functions of RAL estimators for $\psi^{*}$. 

\begin{proposition}\label{Prop: discr-nuisance}For the semiparametric
model characterized by the discrete-time SNMM (\ref{eq:disc-SNMM})
and restriction (\ref{eq:model-part1}), the influence function space
for $\psi^{*}$ is 
\begin{equation}
\Lambda^{\bot}=\left\{ G(\psi^{*};F,c):\ \text{for all }c(\overline{V}_{m})\in\mathbb{\R}^{p}\right\} ,\label{eq:Lambda prop}
\end{equation}
where $\overline{V}_{m}=(\overline{A}_{m-1},\overline{L}_{m})$ and
\[
G(\psi;F,c)=\sum_{m=1}^{K}c(\overline{V}_{m})\{A_{m}-\pr(A_{m}=1\mid\overline{V}_{m})\}[H(\psi)-\E\{H(\psi)\mid\overline{V}_{m}\}],
\]
indexed by $c$. To make the notation accurate, the abbreviation $c$
in $G(\psi;F,c)$ means $c(\overline{V}_{m})$.

\end{proposition}

Although \citet{robins1994correcting} provided this result, the technical
proofs were dense and less accessible to general readers. In the future,
we will write a technical report that provides details to guide general
readers in deriving the semiparametric efficiency theory in similar
contexts.

The semiparametric efficiency score, i.e. the most efficient one among
the class in (\ref{eq:Lambda prop}), often does not have a closed-form
expression. We now make a working assumption, which extends restriction
(\ref{eq:model-part1}) and allows us to derive an analytical expression
of the semiparametric efficient score of $\psi^{*}.$

\begin{assumption}[Homoscedasticity]\label{assump: Homo }For $0\leq m\leq K$,
\textit{ }$\var\{H(\psi^{*})\mid\overline{A}_{m},\overline{L}_{m}\}=\var\left\{ H(\psi^{*})\mid\overline{V}_{m}\right\} $.

\end{assumption}

\begin{proposition}[Discrete-time semiparametric efficient score]\label{Thm: Theorem9}Consider
$\gamma_{m}(\overline{L}_{m};\psi^{*})=(\psi_{1}^{*}+\psi_{2}^{*}t_{m})(\tau-t_{m})$.
Suppose Assumptions \ref{asump:(Consistency)}\textendash \ref{assump: Homo }
holds. The semiparametric efficient score of $\psi^{*}$ is 
\begin{equation}
S_{\eff}(\psi^{*};F)=G(\psi^{*};F,c_{\eff}),\label{eq:semipar ee 1}
\end{equation}
where
\[
c_{\eff}(\overline{V}_{m})=\left(\begin{array}{c}
(\tau-t_{m})-\E\left\{ \dur(t_{m})\mid\overline{A}_{m}=\overline{0},\overline{L}_{m}\right\} \\
t_{m}(\tau-t_{m})-\E\left\{ T\times\dur(t_{m})\mid\overline{A}_{m}=\overline{0},\overline{L}_{m}\right\} 
\end{array}\right)\left[\var\left\{ H(\psi^{*})\mid\overline{V}_{m}\right\} \right]^{-1},
\]
 and $\dur(t_{m})=\sum_{l=m}^{K-1}A_{l}(t_{l+1}-t_{l})$ is the observed
treatment duration from $t_{m}$ to $\tau$.

\end{proposition}

\section{SNMMs in continuous-time processes\label{sec:continuous SNMM}}

\subsection{Setup, models, and assumptions\label{subsec:Setup-cont}}

We now extend the discrete-time SNMM in Section \ref{sec:discrete-SNMM}
to the continuous-time SNMM. We assume that the variables can change
their values at any real time between $0$ and $\tau$. We assume
that all subjects are followed until $\tau$ and consider censoring
in Section \ref{sec:Censoring}.

Each subject has multiple visit times. Let $N(t)$ be the counting
process for the visit times. Let $L_{t}$ be the multidimensional
covariate process. In contrast to the setting with discrete-time data
processes, $L_{t}$ is a vector of covariates at $t$ and additional
information of the past visit times up to but not including $t$.
This is because the past visit pattern, e.g., the number and frequency
of the visit times may be important confounders for the treatment
and outcome processes. Let $A_{t}$ be the binary treatment process.
In our motivating application, the treatment can only be initiated
at the follow-up visits; i.e., if $A_{t}=1$, then $N(t)=1$. We will
model the treatment process directly, although one can model first
the visit time process and then treatment assignment at the visit
times. Define $Y^{(t)}$ as the potential outcome at $\tau$ had the
subject initiated the treatment at $t$, and define $Y^{(\infty)}$
as the potential outcome at $\tau$ had the subject never initiated
the treatment before $\tau$. Let $Y$ be the continuous outcome measured
at $\tau$. For the regularization purpose, we assume that the processes
are Càdlàg processes, i.e., the processes are right continuous with
left limits. Let $V_{t}=(A_{t-},L_{t})$ be the combined treatment
and covariate process, where $A_{t-}$ is the available treatment
information right before $t$. We use the overline notation to denote
a variable's observed history; e.g., $\overline{A}_{t}=\{A_{u}:0\leq u\leq t,\de N(u)=1\}$.
The subject's full record is $F=\{\overline{V}_{\tau},(Y^{(t)}:0\leq t\leq\tau)\}.$
The observed data for a subject through $\tau$ is $O=(\overline{V}_{\tau},Y)$.

We assume the continuous-time SNMM as follows. 

\begin{assumption}[Continuous-time SNMM]\label{cont-SNMMs}For $0\leq t\leq\tau$,
the continuous-time SNMM is
\begin{equation}
\gamma_{t}(\overline{L}_{t})=\E\left\{ Y^{(t)}-Y^{(\infty)}\mid\overline{L}_{t},T\geq t\right\} =\gamma_{t}(\overline{L}_{t};\psi^{*});\label{eq:cont-SNMM}
\end{equation}
i.e., $\gamma_{t}(\overline{L}_{t};\psi)$ with $\psi\in$$\R^{p}$
is a correctly specified model for $\gamma_{t}(\overline{L}_{t})$
with the true parameter value $\psi^{*}.$ Moreover, $Y^{(t)}\sim Y^{(\infty)}+\gamma_{t}(\overline{L}_{t};\psi^{*})$
given $(\overline{L}_{t},T\geq t)$, where $\sim$ means ``is (conditionally)
distributed as''.

\end{assumption}

In the continuous-time SNMM (\ref{eq:cont-SNMM}), $\psi^{*}$ can
be interpreted as the treatment effect rate for the outcome. For the
continuous-time SNMM, we assume that given $(\overline{L}_{t},T\geq t)$,
the treatment effect only changes the location of the distribution
of the outcome but not on other aspects of the distribution such as
the variance. This assumption is stronger than the discrete-time SNMM
in Assumption \ref{asump:disc-SNMMs}. But this assumption is weaker
than the rank-preserving assumption of $Y^{(t)}=Y^{(\infty)}+\gamma_{t}(\overline{L}_{t};\psi^{*})$
considered in \citet{zhang2011causal}. It has been argued that by
mapping the potential outcomes directly rather than between distributions,
rank preserving models are easier to understand and communicate \citep{vansteelandt2014structural}.
However, the rank preservation may be restrictive in practice, because
it implies that for two subjects $i$ and $j$ with the same treatment
and covariate history, $Y_{i}>Y_{j}$ must imply $Y_{i}^{(\infty)}>Y_{j}^{(\infty)}$.
We relax this restriction by imposing a distributional assumption. 

The continuous-time SNMM (\ref{eq:cont-SNMM}) can model the treatment
effect flexibly. For example, the two-parameter model $\gamma_{t}(\overline{L}_{t};\psi^{*})=(\psi_{1}^{*}+\psi_{2}^{*}t)(\tau-t)I(t\leq\tau)$
entails that the treatment effect depends on the treatment initiation
time and the duration of the treatment. To allow for treatment effect
modifiers, we can specify an elaborated treatment effect model including
time-varying covariates, such as viral load in the blood. For example,
one can consider $\gamma_{t}(\overline{L}_{t};\psi^{*})=(\psi_{1}^{*}+\psi_{2}^{*}t+\psi_{3}^{*}\text{lvl}_{t})(\tau-t)I(t\leq\tau)$,
where lvl$_{t}$ is the log viral load at $t$. We discuss effect
modification and model selection in Section \ref{sec:Discussion}.

To link the observed outcome to the potential outcomes, we assume
that $Y=Y^{(T)}$. Define the mimicking outcome for $Y^{(\infty)}$
as $H(\psi^{*})=Y-\gamma_{T}(\overline{L}_{T};\psi^{*})$. By Assumption
\ref{cont-SNMMs}, $H(\psi^{*})\sim Y^{(\infty)},$ given $(\overline{L}_{t},T\geq t)$.

An important issue with data from user-initiated visits and treatment
initiation is the potential selection bias and confounding, e.g.,
sicker patients may visit the doctor more frequently and are likely
to initiate treatment earlier. To overcome this issue, we impose the
NUC assumption on the treatment process \citep{yang2018modeling}.

\begin{assumption}[No unmeasured confounding]\label{asumption:CT-UNC}The
hazard of treatment initiation is 
\begin{eqnarray}
\lambda_{T}(t\mid F) & = & \lim_{h\rightarrow0}h^{-1}P(t\leq T<t+h,\Gamma=1\mid\overline{V}_{t},T\geq t,Y^{(\infty)})\nonumber \\
 & = & \lim_{h\rightarrow0}h^{-1}P(t\leq T<t+h,\Gamma=1\mid\overline{V}_{t},T\geq t)=\lambda_{T}\left(t\mid\overline{V}_{t}\right).\label{eq:UNC}
\end{eqnarray}

\end{assumption}

Assumption \ref{asumption:CT-UNC} implies that the hazard of treatment
initiation at $t$ depends only on the observed treatment and covariate
history $\overline{V}_{t}$ but not on the future observations and
potential outcomes. This assumption holds if the set of historical
covariates contains all prognostic factors for the outcome that affect
the decision of patient visiting the doctor and initiating treatment.
As an example, in the motivating application, time-invariant characteristics
such as age at infection, gender, race and whether ever used injection
drugs are important confounders for the treatment and outcome processes.
Moreover, time-varying CD4 and viral load are important confounders.
Often, poor disease progression necessitates more frequent follow-up
visits and earlier treatment initiation.

The treatment process $A_{t}$ can also be represented in terms of
the counting process $N_{T}(t)$ and the at-risk process $Y_{T}(t)$
of observing treatment initiation. Let $\sigma(V_{t})$ be the $\sigma$-field
generated by $V_{t}$, and let $\sigma(\overline{V}_{t})$ be the
$\sigma$-field generated by $\cup_{u\leq t}\sigma(V_{u})$. Under
the standard regularity conditions for the counting process, $M_{T}(t)=N_{T}(t)-\int_{0}^{t}\lambda_{T}(u\mid\overline{V}_{u})Y_{T}(u)\de u$
is a martingale with respect to the filtration $\sigma(\overline{V}_{t})$.
Assumption \ref{asumption:CT-UNC} entails that the jumping rate of
$N_{T}(t)$ at $t$ does not depend on $Y^{(\infty)}$, given $\overline{V}_{t}$.
Because $H(\psi^{*})$ mimics $Y^{(\infty)}$ in the sense that it
has the same distribution as $Y^{(\infty)}$ given $\overline{V}_{t}$,
Assumption \ref{asumption:CT-UNC} also implies that the jumping rate
of $N_{T}(t)$ at $t$ does not depend on $H(\psi^{*})$, given $\overline{V}_{t}$.
To be formal, we show in the supplementary material that
\begin{equation}
\lambda_{T}\{t\mid\overline{V}_{t},H(\psi^{*})\}=\lambda_{T}(t\mid\overline{V}_{t}).\label{eq:UNC2}
\end{equation}
Therefore, under the standard regularity conditions, $M_{T}(t)$ is
a martingale with respect to the filtration $\sigma\{\overline{V}_{t},H(\psi^{*})\}$.
\citet{lok2008statistical} imposed this martingale condition to formulate
the NUC assumption for the treatment process. 

\subsection{Semiparametric efficiency score}

To estimate the causal parameter precisely, we establish the new semiparametric
efficiency theory for the continuous-time SNMMs. We defer all proofs
to the supplementary material.

\begin{theorem}\label{Thm: cont-nuisance}For the semiparametric
model characterized by the continuous-time SNMM (\ref{eq:cont-SNMM})
and Assumption \ref{asumption:CT-UNC}, the influence function space
for $\psi^{*}$ is 
\[
\Lambda^{\bot}=\left\{ G(\psi^{*};F,c):\ \text{for all }c(\overline{V}_{u})\in\mathbb{\R}^{p}\right\} ,
\]
where 
\begin{equation}
G(\psi;F,c)=\int_{0}^{\tau}c(\overline{V}_{u})\left[H(\psi)-\E\left\{ H(\psi)\mid\overline{V}_{u},T\geq u\right\} \right]Y_{T}(u)\de M_{T}(u).\label{eq:G}
\end{equation}

\end{theorem}

The semiparametric efficiency score for $\psi^{*}$ is $S_{\eff}(\psi^{*};F)=\prod\{S(\psi^{*};F)\mid\Lambda^{\bot}\}$.
To derive $S_{\eff}(\psi^{*};F)$, we calculate the projection of
any $B=B(F)$ onto $\Lambda^{\bot}.$

\begin{theorem}\label{Thm:projection}For any $B=B(F)$, the projection
of $B$ onto $\Lambda^{\bot}$ is 
\begin{multline}
\prod\left(B\mid\Lambda^{\bot}\right)=\int_{0}^{\tau}\left[\E\left\{ B\dot{H}_{u}(\psi^{*})\mid\overline{V}_{u},T=u\right\} -\E\left\{ B\dot{H}_{u}(\psi^{*})\mid\overline{V}_{u},T\geq u\right\} \right]\\
\times\left[\var\left\{ H(\psi^{*})\mid\overline{V}_{u},T\geq u\right\} \right]^{-1}\left[H(\psi^{*})-\E\left\{ H(\psi^{*})\mid\overline{V}_{u},T\geq u\right\} \right]\de M_{T}(u),\label{eq:projection}
\end{multline}
where $\dot{H}_{u}(\psi^{*})=H(\psi^{*})-\E\{H(\psi^{*})\mid\overline{V}_{u},T\geq u\}$.

\end{theorem}

Considering $B=S(\psi^{*};F)$ in Theorem \ref{Thm:projection}, we
can derive the semiparametric efficient score for $\psi^{*}$.

\begin{theorem}[Continuous-time semiparametric efficient score]\label{Thm: efficient score}For
the semiparametric model characterized by the continuous-time SNMM
(\ref{eq:cont-SNMM}) and Assumption \ref{asumption:CT-UNC}, the
semiparametric efficient score of $\psi^{*}$ is 
\begin{equation}
S_{\mathrm{eff}}(\psi^{*};F)=G(\psi^{*};F,c_{\mathrm{eff}}),\label{eq:semipar score}
\end{equation}
where $G(\psi;F,c)$ is defined in (\ref{eq:G}), and 
\begin{multline}
c_{\mathrm{eff}}(\overline{V}_{u})=[\E\{\partial\dot{H}_{u}(\psi^{*})/\partial\psi\mid\overline{V}_{u},T=u\}\\
-\E\{\partial\dot{H}_{u}(\psi^{*})/\partial\psi\mid\overline{V}_{u},T\geq u\}]\times[\var\{H(\psi^{*})\mid\overline{V}_{u},T\geq u\}]^{-1}.\label{eq:c_eff}
\end{multline}

\end{theorem}

To illustrate the theorem, we provide the explicit expression of the
semiparametric efficient score using an example.

\begin{example}Consider $\gamma_{t}(\overline{L}_{t};\psi)=(\psi_{1}+\psi_{2}t)(\tau-t)I(t\leq\tau)$.
Suppose Assumption \ref{asumption:CT-UNC} holds. The semiparametric
efficient score of $\psi^{*}$ is $S_{\mathrm{eff}}(\psi^{*};F)=G(\psi^{*};F,c_{\mathrm{eff}})$,
where
\begin{multline}
c_{\mathrm{eff}}(\overline{V}_{u})=\left(\begin{array}{c}
(\tau-u)I(u\leq\tau)-\E\{(\tau-T)I(T\leq\tau)\mid\overline{V}_{u},T\geq u\}\\
u(\tau-u)I(u\leq\tau)-\E\{T(\tau-T)I(T\leq\tau)\mid\overline{V}_{u},T\geq u\}
\end{array}\right)\\
\times[\var\{H(\psi^{*})\mid\overline{V}_{u},T\geq u\}]^{-1}.\label{eq:c_eff_eg}
\end{multline}

\end{example}

\begin{remark}

The proposed continuous-time semiparametric efficient score contains
the discrete-time semiparametric efficient score as a special case.
If the processes take observations at discrete times $\{t_{0},\ldots,t_{K}\}$,
then (i) the conditioning event $(\overline{V}_{u},T\geq u)$ at $t_{m}$
is the same as $(\overline{A}_{m}=\overline{0},\overline{L}_{m})$,
(ii) $M_{T}(t)=N_{T}(t)-\int_{0}^{t}\lambda_{T}(u\mid\overline{V}_{u})Y_{T}(u)\de u$
at $t=t_{m}$ becomes $A_{m}-\pr(A_{m}=1\mid\overline{A}_{m-1}=\overline{0},\overline{L}_{m})$,
and $\E\{\partial\dot{H}_{t}(\psi^{*})/\partial\psi\mid\overline{V}_{t},T=t\}$
at $t=t_{m}$ becomes
\[
\E\{\partial\dot{H}_{m}(\psi^{*})/\partial\psi\mid\overline{V}_{m},T=t_{m}\}=-\left(\begin{array}{c}
(\tau-t_{m})-\E\left\{ \dur(t_{m})\mid\overline{A}_{m}=\overline{0},\overline{L}_{m}\right\} \\
t_{m}(\tau-t_{m})-\E\left\{ T\times\dur(t_{m})\mid\overline{A}_{m}=\overline{0},\overline{L}_{m}\right\} 
\end{array}\right).
\]
Therefore, the continuous-time semiparametric efficient score (\ref{eq:semipar score})
reduces to the discrete-time semiparametric efficient score (\ref{eq:semipar ee 1}).

\end{remark}

\subsection{Doubly robust and locally efficient estimators}

We now construct a general class of estimators based on the estimating
function $G(\psi^{*};F,c)$. Because $\E\{G(\psi^{*};F,c)\}=0$, we
obtain the estimator of $\psi^{*}$ by solving 
\begin{equation}
\bP_{n}\left\{ G(\psi;F,c)\right\} =0.\label{eq:ee4}
\end{equation}
In particular, the estimating equation (\ref{eq:ee4}) with $c_{\eff}$
provides the semiparametric efficient estimator of $\psi^{*}$.

In (\ref{eq:ee4}), we assume that the model for the treatment process
and $\E\left\{ H(\psi^{*})\mid\overline{V}_{u},T\geq u\right\} $
are known. In practice, they are often unknown and must be modeled
and estimated from the data. We posit a proportional hazards model
with time-dependent covariates for the treatment process; i.e., 
\begin{eqnarray}
\lambda_{T}\left(t\mid\overline{V}_{t};\alpha\right) & = & \lambda_{T,0}(t)\exp\left\{ \alpha^{\T}W_{T}(t,\overline{V}_{t})\right\} ,\label{eq:ph-V}
\end{eqnarray}
where $\lambda_{T,0}(t)$ is an unknown baseline hazard function,
$W_{T}(t,\overline{V}_{t})$ is a pre-specified function of $t$ and
$\overline{V}_{t}$, and $\alpha$ is a vector of unknown parameters.
Under Assumption \ref{asumption:CT-UNC}, we can estimate $\lambda_{T,0}(t)$
and $\alpha$ from the standard software such as ``coxph'' in R
(R Development Core Team, 2012) \nocite{R:2010}. To estimate $\alpha$,
fit the time-dependent proportional hazards model to the data $\{(\overline{V}_{T_{i},i},T_{i},\Gamma_{i}):i=1,\ldots,n\}$
treating the treatment initiation as the failure event. Once we obtain
$\widehat{\alpha},$ we can estimate the cumulative baseline hazard,
$\lambda_{T,0}(t)\de t$ by
\[
\widehat{\lambda}_{T,0}(t)\de t=\frac{\sum_{i=1}^{n}\de N_{T,i}(t)}{\sum_{i=1}^{n}\exp\left\{ \widehat{\alpha}^{\T}W_{T}(t,\overline{V}_{t,i})\right\} Y_{T_{i}}(t)}.
\]
Then, we obtain $\widehat{\lambda}_{T}(u\mid\overline{V}_{u})=\exp\left\{ \widehat{\alpha}^{\T}W_{T}(u,\overline{V}_{u})\right\} \widehat{\lambda}_{T,0}(u)$
and $\widehat{M}_{T}(t)=N_{T}(t)-\int_{0}^{t}\widehat{\lambda}_{T}(u\mid\overline{V}_{u})Y_{T}(u)\de u$. 

We also posit a working model $\E\left\{ H(\psi^{*})\mid\overline{V}_{u},T\geq u;\beta\right\} $,
such as a linear regression model, where $\beta$ is a vector of unknown
parameters.

The estimating equation for $\psi^{*}$ achieves the double robustness
or double protection \citep{rotnitzky2015double}.

\begin{theorem}[Double robustness]\label{Thm:2-dr}Under the continuous-time
SNMM (\ref{eq:cont-SNMM}) and Assumption \ref{asumption:CT-UNC},
the proposed estimator $\widehat{\psi}$ solving the estimating equation
(\ref{eq:ee4}) is doubly robust in that it is unbiased if either
the model for the treatment process is correctly specified, or the
potential outcome mean model $\E\left\{ H(\psi^{*})\mid\overline{V}_{u},T\geq u;\beta\right\} $
is correctly specified, but not necessarily both.

\end{theorem}

The choice of $c$ does not affect the double robustness but the efficiency
of the resulting estimator. For efficiency consideration, we consider
$c_{\eff}$ in (\ref{eq:c_eff}). The resulting estimator solving
the estimating equation (\ref{eq:ee4}) with $c_{\eff}$ is locally
efficient, in the sense that it achieves the semiparametric efficiency
bound if the working models for the treatment process and the potential
outcome mean are correctly specified. Because $c_{\eff}$ depends
on the unknown distribution, we require additional models for $\E\{(\tau-T)I(T\leq\tau)\mid\overline{V}_{u},T\geq u\}$
and $\E\{T(\tau-T)I(T\leq\tau)\mid\overline{V}_{u},T\geq u\}$ to
approximate $c_{\eff}.$ For example, we can approximate $\E\{(\tau-T)I(T\leq\tau)\mid\overline{V}_{u},T\geq u\}$
by $P(T\leq\tau\mid\overline{V}_{u},T\geq u)\times\E\{\tau-T\mid\overline{V}_{u},u\leq T\leq\tau\}$
and each approximated by (logistic) linear models. For $\mathrm{\var}\{H(\psi^{*})\mid\overline{V}_{u},T\geq u\}$,
we consider the following options: (i) assume $\var\{H(\psi^{*})\mid\overline{V}_{u},T\geq u\}$
to be a constant, and (ii) approximate $\var\{H(\psi^{*})\mid\overline{V}_{u},T\geq u\}$
by the sample variance of $H(\widehat{\psi}_{p})$ among subjects
with $T\geq u$, where $\widehat{\psi}_{p}$ is a preliminary estimator.
We compare the two options via simulation. Although option (ii) provides
a slight efficiency gain in estimation, for ease of implementation
we recommend option (i). Option (i) is common in the generalized estimating
equation framework. From here on, we use this option for $c$ and
suppress the dependence on $c$ for estimating functions.

\subsection{Censoring\label{sec:Censoring}}

As in the AIEDRP study, in most longitudinal observational studies,
subjects may drop out the study prematurely before the end of study,
which renders the data censored at the time of dropout. If the censoring
mechanism depends on time-varying prognostic factors, e.g. sicker
patients drop out of the study with a higher probability than healthier
patients, the patients remaining in the study is a biased sample of
the full population. We now introduce $C$ to be the time to censoring.
Let $X=\min(C,\tau)$ be time to censoring or the end of the study,
whichever came first. Let $\delta_{C}=I(C\geq\tau)$ be the indicator
of not censoring before $\tau$. The observed data is $O=(X,\overline{V}_{X},\delta_{C},\delta_{C}Y)$. 

In the presence of censoring, the estimating equation (\ref{eq:ee4})
is not feasible. We consider inverse probability of censoring weighting
(IPCW; \citealp{robins1993information}). We assume a dependent censoring
mechanism as follows.

\begin{assumption}[Dependent censoring]\label{asp:NUC-1}The hazard
of censoring is 
\begin{eqnarray}
\lambda_{C}(t\mid F,T>t) & = & \lim_{h\rightarrow0}h^{-1}P(t\leq C<t+h\mid F,T>t,C\geq t)\nonumber \\
 & = & \lim_{h\rightarrow0}h^{-1}P(t\leq C<t+h\mid\overline{V}_{t},T>t,C\geq t)=\lambda_{C}\left(t\mid\overline{V}_{t}\right).\label{eq:censoring}
\end{eqnarray}

\end{assumption}

Assumption \ref{asp:NUC-1} states that $\lambda_{C}(t\mid F,T>t)$
depends only on the past treatment and covariate history until $t$,
but not on the future variables and potential outcomes. This assumption
holds if the set of historical covariates contains all prognostic
factors for the outcome that affect the possibility of loss to follow
up at $t$. Under this assumption, the missing data due to censoring
are missing at random \citep{rubin1976inference}. 

We discuss the implication of Assumption \ref{asp:NUC-1} on estimation
of the treatment process model. Under Assumption \ref{asp:NUC-1},
the hazard of treatment initiation in (\ref{eq:UNC}) is equal to
$\lim_{h\rightarrow0}h^{-1}P(t\leq T<t+h,\Gamma=1\mid\overline{V}_{t},T>t,C\geq t)$.
Redefining $T$ to be the time to treatment initiation, or censoring,
or the end of the study, whichever came first, (\ref{eq:UNC}) can
be estimated by conditioning on $T\geq t$ with the new definition
of $T.$ 

From $\lambda_{C}\left(t\mid\overline{V}_{t}\right)$, we define $K_{C}\left(t\mid\overline{V}_{t}\right)=\exp\left\{ -\int_{0}^{t}\lambda_{C}\left(u\mid\overline{V}_{u}\right)\de u\right\} ,$
which is the probability of the subject not being censored before
$t$. For regularity, we impose a positivity condition for $K_{C}\left(t\mid\overline{V}_{t}\right)$.

\begin{assumption}[Positivity]\label{asp:positivity}There exists
a constant $\delta$ such that with probability one, $K_{C}\left(t\mid\overline{V}_{t}\right)\geq\delta>0$
for $t$ in the support of $T$.

\end{assumption}

Following \citet{rotnitzky2007analysis}, we obtain the IPCW estimator
$\widehat{\psi}$ as the solution to the following equation: 
\begin{equation}
\bP_{n}\left\{ \frac{\delta_{C}}{K_{C}(\tau\mid\overline{V}_{\tau})}G(\psi;F)\right\} =0.\label{eq:IPCW ee}
\end{equation}

In (\ref{eq:IPCW ee}), we assume that $K_{C}(t\mid\overline{V}_{t})$
is known. In practice, $K_{C}(t\mid\overline{V}_{t})$ is often unknown
and must be modeled and estimated from the data. To facilitate estimation,
we posit a proportional hazards model for the censoring process with
time-dependent covariates; i.e., 
\begin{equation}
\lambda_{C}(t\mid\overline{V}_{t})=\lambda_{C,0}(t)\exp\{\eta{}^{\T}W_{C}(t,\overline{V}_{t})\},\label{eq:ph-C}
\end{equation}
where $\lambda_{C,0}(t)$ is an unknown baseline hazard function for
censoring, $W_{C}(t,\overline{V}_{t})$ is a pre-specified function
of $t$ and $\overline{V}_{t}$, and $\eta$ is a vector of unknown
parameters. Under Assumption \ref{asp:NUC-1}, we can estimate $\lambda_{C,0}(t)$
and $\alpha$ from the standard software such as ``coxph'' in R.
To estimate $\eta$, fit the time-dependent proportional hazards model
to the data $\{(\overline{V}_{X_{i},i},X_{i},\delta_{C,i}):i=1,\ldots,n\}$
treating the censoring as the failure event. Once we obtain $\widehat{\eta},$
we can estimate $\lambda_{C,0}(t)\de t$ by
\[
\widehat{\lambda}_{C,0}(t)\de t=\frac{\sum_{i=1}^{n}\de N_{C,i}(t)}{\sum_{i=1}^{n}\exp\left\{ \widehat{\eta}^{\T}W_{C}(t,\overline{V}_{t,i})\right\} Y_{C_{i}}(t)},
\]
where $N_{C}(t)=I(C\leq t,\delta_{C}=0)$ and $Y_{C}(t)=I(C\geq t)$
are the counting process and the at-risk process of observing censoring.
Then, we estimate $K_{C}\left(t\mid\overline{V}_{t}\right)$ by 
\begin{eqnarray*}
\widehat{K}_{C}\left(t\mid\overline{V}_{t}\right) & = & \exp\left[-\int_{0}^{t}\exp\{\widehat{\eta}^{\T}W_{C}(u,\overline{V}_{u})\}\widehat{\lambda}_{C,0}(u)\de u\right]\\
 & = & \prod_{0\leq u\leq t}\left[1-\exp\left\{ \widehat{\eta}^{\T}W_{C}(u,\overline{V}_{u})\right\} \widehat{\lambda}_{C,0}\left(u\right)\de u\right].
\end{eqnarray*}
Then, we obtain the estimator $\widehat{\psi}$ of $\psi$ by solving
(\ref{eq:IPCW ee}) with unknown quantities replaced by their estimates.

In the literature, augmented IPCW estimators have been developed to
improve efficiency and robustness over IPCW estimators; see, e.g.,
\citet{rotnitzky2007analysis,rotnitzky2009analysis} for survival
data and \citet{lok2016cumincidentfunction} for competing risks data.
However, the typical efficiency gain is little in practice at the
expense of additional complexity in computation. More importantly,
we show in the next section that the proposed IPCW estimator already
has the multiple robustness property against possible model misspecification.

\section{Multiple robustness and asymptotic distribution\label{sec:Asymptotic-property}}

Because the proposed estimator depends on nuisance parameter estimation,
we summarize the following nuisance models: (i) $\E\{H(\psi^{*})\mid\overline{V}_{u},T\geq u;\beta\}$
indexed by $\beta$; (ii) the proportional hazards model for the treatment
process (\ref{eq:ph-V}), denoted by $M_{T}$; and (iii) the proportional
hazards model for the censoring process (\ref{eq:ph-C}), denoted
by $K_{C}$. Let $\widehat{\beta}$, $\widehat{M}_{T}$, and $\widehat{K}_{C}$
be the estimates of $\beta$, $M_{T}$, and $K_{C}$ under the specified
parametric and semiparametric models. Denote the probability limits
of $\widehat{\beta}$, $\widehat{M}_{T}$, and $\widehat{K}_{C}$
as $\beta^{*}$, $M_{T}^{*}$, and $K_{C}^{*}$, respectively. If
the outcome model is correctly specified, $\E\{H(\psi^{*})\mid\overline{V}_{u},T\geq u;\beta^{*}\}=\E\{H(\psi^{*})\mid\overline{V}_{u},T\geq u\}$;
if the model for the treatment process is correctly specified, $M_{T}^{*}=M_{T}$;
and if the model for the censoring process is correctly specified,
$K_{C}^{*}=K_{C}$. To reflect that the estimating function depends
on the nuisance parameters, we denote
\begin{eqnarray*}
G(\psi,\beta,M_{T};F) & = & \int c(\overline{V}_{u})\left[H(\psi)-\E\left\{ H(\psi)\mid\overline{V}_{u},T\geq u;\beta\right\} \right]\de M_{T}(u),\\
\Phi(\psi,\beta,M_{T},K_{C};F) & = & \frac{\delta_{C}G(\psi,\beta,\lambda_{T};F)}{K_{C}\left(\tau\mid\overline{V}_{\tau}\right)}.
\end{eqnarray*}
Then, the proposed estimator $\widehat{\psi}$ solves
\begin{equation}
\bP_{n}\left\{ \Phi(\psi,\widehat{\beta},\widehat{M}_{T},\widehat{K}_{C};F)\right\} =0,\label{eq:IPCW2-1}
\end{equation}
for $\psi$, which achieves the multiple robustness or multiple protection
\citep{molina2017multiple}.

\begin{theorem}[Multiple robustness]\label{Thm:3-mr}Under the continuous-time
SNMM (\ref{eq:cont-SNMM}) and Assumption \ref{asumption:CT-UNC},
the proposed estimator $\widehat{\psi}$ solving estimating equation
(\ref{eq:IPCW2-1}) is multiply robust in that it is unbiased under
all scenarios specified in Table \ref{tab:Multiply-Robustness}.

\begin{table}[h]
\protect\protect\protect\protect\protect\caption{\label{tab:Multiply-Robustness}Multiply Robustness of the Proposed
Estimator}

\centering{}%
\begin{tabular}{llccccccccccc}
\hline 
\multicolumn{13}{l}{The proposed estimator $\widehat{\psi}$ is unbiased if}\tabularnewline
\hline 
(i) Model for $H(\psi^{*})$ &  & $\checked$ &  & $\checked$ &  & $\times$ &  & $\checked$ &  & $\checked$ &  & $\times$\tabularnewline
(ii) Model for the treatment process $M_{T}$ &  & $\checked$ &  & $\times$ &  & $\checked$ &  & $\checked$ &  & $\times$ &  & $\checked$\tabularnewline
(iii) Model for the censoring process $K_{C}$ &  & $\checked$ &  & $\checked$ &  & $\checked$ &  & $\times$ &  & $\times$ &  & $\times$\tabularnewline
\hline 
\end{tabular}

$\checked$ (is correctly specified), $\times$ (is misspecified)
\end{table}

\end{theorem}

It is important to establish the asymptotic property of $\widehat{\psi}$
under the multiple robustness condition, which allows for multiply
robust inference of $\psi^{*}$. Let $P$ denote the true data generating
distribution of $F$, and for any $g(F)$, let $\bP\{g(F)\}=\int g(f)\de P(f)$
and let $\G_{n}=n^{1/2}(\bP_{n}-\bP)$. We define 
\begin{eqnarray*}
J_{1}(\beta) & = & \bP\left\{ \Phi(\psi^{*},\beta,M_{T}^{*},K_{C}^{*};F)\right\} ,\\
J_{2}(M_{T}) & = & \bP\left\{ \Phi(\psi^{*},\beta^{*},M_{T},K_{C}^{*};F)\right\} ,\\
J_{3}(K_{C}) & = & \bP\left\{ \Phi(\psi^{*},\beta^{*},M_{T}^{*},K_{C};F)\right\} ,
\end{eqnarray*}
and 
\[
J(\beta,M_{T},K_{C})=\bP\left\{ \Phi(\psi^{*},\beta,M_{T},K_{C};F)\right\} .
\]

Similar to \citet{yang2015gof}, we impose the regularity conditions
from the empirical process literature \citep{van1996weak}.

\begin{assumption}\label{asump:donsker}
\begin{description}
\item [{(i)}] $\Phi(\psi,\beta,M_{T},K_{C};F)$ and $\partial\Phi(\psi,\beta,M_{T},K_{C};F)/\partial\psi$
are $P$-Donsker classes; i.e.,
\begin{eqnarray*}
\G_{n}\{\Phi(\widehat{\psi},\widehat{\beta},\widehat{M}_{T},\widehat{K}_{C};F)\} & = & \G_{n}\{\Phi(\psi^{*},\beta^{*},M_{T}^{*},K_{C}^{*};F)\}+o_{p}(1),\\
\G_{n}\left\{ \frac{\partial\Phi(\widehat{\psi},\widehat{\beta},\widehat{M}_{T},\widehat{K}_{C};F)}{\partial\psi}\right\}  & = & \G_{n}\left\{ \frac{\partial\Phi(\psi^{*},\beta^{*},M_{T}^{*},K_{C}^{*};F)}{\partial\psi}\right\} +o_{p}(1).
\end{eqnarray*}
\item [{(ii)}] Assume that
\begin{eqnarray*}
\bP\left\{ ||\Phi(\psi^{*},\widehat{\beta},\widehat{M}_{T},\widehat{K}_{C};F)-\Phi(\psi^{*},\beta^{*},M_{T}^{*},K_{C}^{*};F)||\right\}  & = & o_{p}(1),\\
\bP\left\{ ||\frac{\partial}{\partial\psi}\Phi(\widehat{\psi},\widehat{\beta},\widehat{M}_{T},\widehat{K}_{C};F)-\frac{\partial}{\partial\psi}\Phi(\psi^{*},\beta^{*},M_{T}^{*},K_{C}^{*};F)||\right\}  & = & o_{p}(1).
\end{eqnarray*}
\item [{(iii)}] $A(\psi^{*},\beta^{*},M_{T}^{*},K_{C}^{*})=\bP\left\{ \partial\Phi(\psi^{*},\beta^{*},M_{T}^{*},K_{C}^{*};F)/\partial\psi\right\} $
is invertible.
\item [{(iv)}] Assume that 
\begin{multline*}
J(\widehat{\beta},\widehat{M}_{T},\widehat{K}_{C})-J(\beta^{*},M_{T}^{*},K_{C}^{*})=J_{1}(\widehat{\beta})-J_{1}(\beta^{*})+J_{2}(\widehat{M}_{T})-J_{2}(M_{T}^{*})\\
+J_{3}(\widehat{K}_{C})-J_{3}(K_{C}^{*})+o_{p}(n^{-1/2}),
\end{multline*}
and that $J_{1}(\widehat{\beta})$, $J_{2}(\widehat{M}_{T})$, and
$J_{3}(\widehat{K}_{C})$ are regular asymptotically linear with influence
functions $\Phi_{1}(\psi^{*},\beta^{*},M_{T}^{*},K_{C}^{*};F)$, $\Phi_{2}(\psi^{*},\beta^{*},M_{T}^{*},K_{C}^{*};F)$,
and $\Phi_{3}(\psi^{*},\beta^{*},M_{T}^{*},K_{C}^{*};F)$, respectively.
\end{description}
\end{assumption}

We discuss the implications of these conditions. First, the $P$-Donsker
class condition requires that the nuisance models should not be too
complex. Under Assumption \ref{asp:positivity} for the censoring
process, Assumption \ref{asump:donsker} (i) is a standard condition
for the empirical processes. We refer the interested readers to Section
4.2 of \citet{kennedy2016semiparametric} for a thorough discussion
of Donsker classes of functions. Second, Assumption \ref{asump:donsker}
(ii) states that $\widehat{\beta}$, $\widehat{M}_{T}$, and $\widehat{K}_{C}$
have probability limits $\beta^{*}$, $M_{T}^{*}$, and $K_{C}^{*}$,
and that the multiple robustness condition in Theorem \ref{Thm:3-mr}
holds. Third, Assumption \ref{asump:donsker} (iv) holds for smooth
functionals of parametric or semiparametric efficient estimators under
specified models. Therefore, this assumption would hold under mild
regularity conditions if $\widehat{\beta}$, $\widehat{M}_{T}$, and
$\widehat{K}_{C}$ are the parametric and semiparametric maximum likelihood
estimators under specified models.

We present the asymptotic property of the proposed estimator $\widehat{\psi}$
solving equation (\ref{eq:IPCW2-1}).

\begin{theorem}\label{thm:4}Under the continuous-time SNMM (\ref{eq:cont-SNMM})
and Assumptions \ref{asumption:CT-UNC}, \ref{asp:positivity} and
\ref{asump:donsker}, $\widehat{\psi}$ is consistent for $\psi^{*}$
and is asymptotically linear with the influence function 
\[
\widetilde{\Phi}(\psi^{*},\beta^{*},M_{T}^{*},K_{C}^{*};F)=\left\{ A(\psi^{*},\beta^{*},M_{T}^{*},K_{C}^{*})\right\} ^{-1}\widetilde{B}(\psi^{*},\beta^{*},M_{T}^{*},K_{C}^{*};F),
\]
where $A(\psi^{*},\beta^{*},M_{T}^{*},K_{C}^{*})$ is defined in Assumption
\ref{asump:donsker} (iii), and 
\begin{eqnarray}
\widetilde{B}(\psi^{*},\beta^{*},M_{T}^{*},K_{C}^{*};F) & = & \Phi(\psi^{*},\beta^{*},M_{T}^{*},K_{C}^{*};F)+\Phi_{1}(\psi^{*},\beta^{*},M_{T}^{*},K_{C}^{*};F)\nonumber \\
 &  & +\Phi_{2}(\psi^{*},\beta^{*},M_{T}^{*},K_{C}^{*};F)+\Phi_{3}(\psi^{*},\beta^{*},M_{T}^{*},K_{C}^{*};F).\label{eq:influence fctn}
\end{eqnarray}

\end{theorem}

Theorem \ref{thm:4} allows for variance estimation of $\widehat{\psi}$.
If the nuisance models are correctly specified, we have
\begin{eqnarray}
\widetilde{B}(\psi^{*},\beta^{*},M_{T},K_{C};F) & = & \Phi(\psi^{*},\beta^{*},M_{T},K_{C};F)-\E\left\{ \Phi(\psi^{*},\beta^{*},M_{T},K_{C};F)S_{\alpha}^{\T}\right\} \E\left(S_{\alpha}S_{\alpha}^{\T}\right)^{-1}S_{\alpha}\nonumber \\
 &  & -\E\left\{ \Phi(\psi^{*},\beta^{*},M_{T},K_{C};F)S_{\eta}^{\T}\right\} \E\left(S_{\eta}S_{\eta}^{\T}\right)^{-1}S_{\eta}\nonumber \\
 &  & +\int\frac{\E\left[G(\psi^{*},\beta^{*},M_{T};F)\exp\left\{ \alpha^{\T}W_{T}(u,\overline{V}_{u})\right\} \delta_{C}/K_{C}(\tau\mid\overline{V}_{\tau})\right]}{\E\left[\exp\left\{ \alpha^{\T}W_{T}(u,\overline{V}_{u})\right\} Y_{T}(u)\right]}\de M_{C}(u)\nonumber \\
 &  & +\int\frac{\E\left[G(\psi^{*},\beta^{*},M_{T};F)\exp\left\{ \eta^{\T}W_{C}(u,\overline{V}_{u})\right\} \delta_{C}/K_{C}(\tau\mid\overline{V}_{\tau})\right]}{\E\left[\exp\left\{ \eta^{\T}W_{C}(u,\overline{V}_{u})\right\} Y_{C}(u)\right]}\de M_{T}(u),\label{eq:tilde-J-1}
\end{eqnarray}
where $S_{\alpha}$ and $S_{\eta}$ are the scores of the partial
likelihood functions of $\alpha$ and $\eta$, respectively; see (\ref{eq:S_alpha})
and (\ref{eq:S_eta}) in the supplementary material.

Then, we obtain the variance estimator of $\widehat{\psi}$ as the
empirical variance of the individual influence function with the unknown
parameters replaced by their estimates. Under the multiple robustness
condition if some nuisance models are misspecified, it is difficult
to characterize the influence function $\widetilde{\Phi}(\psi^{*},\beta^{*},M_{T}^{*},K_{C}^{*};F)$.
We suggest estimating the asymptotic variance of $\widehat{\psi}$
by nonparametric bootstrap \citep{efron1979}. The consistency of
the bootstrap is guaranteed by the regularity and asymptotic properties
of $\widehat{\psi}$ in Theorem \ref{thm:4}.

\section{Simulation study\label{sec:simulation}}

We now evaluate the finite-sample performance of the proposed estimator
on simulated datasets with two objectives. First, we assess the double
robustness and efficiency of the proposed estimator based on the semiparametric
efficiency score, compared to some preliminary estimator. Second,
to demonstrate the impact of data discretization as commonly done
in practice, we include the g-estimator applied to the pre-processed
data.

We simulate $1,000$ datasets under two settings with and without
censoring. In Setting I, we generate two covariates, one time-independent
($L_{TI}$) and one time-dependent ($L_{TD}$). The time-independent
covariate $L_{TI}$ is generated from a Bernoulli distribution with
mean equal to $0.55$. The time-dependent covariate is $L_{TD,t}=l_{1}\times I(0\leq t<0.5)+l_{2}\times I(0.5\leq t<1)+l_{3}\times I(1\leq t<1.5)+l_{4}\times I(1.5\leq t\leq2)$,
where $(l_{1},l_{2},l_{3},l_{4})^{\T}$ is a $1\times4$ row vector
generated from a multivariate normal distribution with mean equal
to $(0,0,0,0)$ and covariance equal to $0.7^{|i-j|}$ for$i,j=1,\ldots,4$.
We assume that the time-dependent variable remains constant between
measurements. The maximum follow up time is $\tau=2$ (in year). We
generate the time to treatment initiation $T$ with the hazard rate
$\lambda_{T}(t\mid\overline{V}_{t})=\lambda_{T,0}(t)\exp($$\alpha_{1}$
$\times L_{TI}+\alpha_{2}L_{TD,t})$ with $\lambda_{T,0}(t)=\lambda_{T,0}=0.4$,
$\alpha_{1}=0.15$, and $\alpha_{2}=0.8$. We generate $T$ according
to the time-dependent model sequentially. This is because the hazard
of treatment initiation in the time interval from $t_{1}=0$ to $t_{2}=0.5$
differs from the hazard of treatment initiation in the next interval
and so on; see the supplementary material for details. We let $Y^{(\infty)}=L_{TD,\tau}$
be the potential outcome had the subject never initiated the treatment
before $\tau$. The observed outcome is $Y=Y^{(\infty)}+\gamma_{T}(\overline{V}_{T};\psi^{*})$,
where $\gamma_{t}(\overline{V}_{t};\psi^{*})=(\psi_{1}^{*}+\psi_{2}^{*}t)(\tau-t)I(t\leq\tau)$
with $\psi_{1}^{*}=15$ and $\psi_{2}^{*}=-1$.
\begin{table}
\caption{\label{tab:results1}Simulation results in Setting I without censoring
based on $1,000$ simulated datasets: the Monte Carlo bias, standard
error, root mean square error of the estimators, and coverage rate
of $95\%$ confidence intervals.}

\centering{}%
\begin{tabular}{cclcccccccc}
\hline 
 &  &  & \multicolumn{2}{c}{Bias ($\times10^{2}$)} & \multicolumn{2}{c}{SE ($\times10^{2}$)} & \multicolumn{2}{c}{rMSE ($\times10^{2}$)} & \multicolumn{2}{c}{CR ($\times10^{2}$)}\tabularnewline
$n$ &  & Method & $\psi_{1}^{*}$ & $\psi_{2}^{*}$ & $\psi_{1}^{*}$ & $\psi_{2}^{*}$ & $\psi_{1}^{*}$ & $\psi_{2}^{*}$ & $\psi_{1}^{*}$ & $\psi_{2}^{*}$\tabularnewline
\hline 
\multicolumn{11}{c}{Scenario (i) with $M_{T}$ ($\checked$)}\tabularnewline
\hline 
 & Model for $H(\psi^{*})$ $(\times)$ & $\widehat{\psi}_{p}$ & 0.3 & -0.1 & 5.3 & 9.6 & 5.3 & 9.6 & 95.0 & 94.0\tabularnewline
\multirow{2}{*}{$1000$} & \multirow{2}{*}{Model for $H(\psi^{*})$ $(\checked)$} & $\widehat{\psi}_{\cont,1}$ & 0.2 & 0.1 & 5.0 & 8.9 & 5.0 & 8.9 & 95.4 & 94.0\tabularnewline
 &  & $\widehat{\psi}_{\cont,2}$ & 0.2 & 0.1 & 4.9 & 8.7 & 4.9 & 8.7 & 95.3 & 94.4\tabularnewline
 & \textendash{} & $\widehat{\psi}_{\disc,g}$ & 28.6 & 34.5 & 6.0 & 10.5 & 29.3 & 36.1 & 0.0 & 7.2\tabularnewline
 & Model for $H(\psi^{*})$ $(\times)$ & $\widehat{\psi}_{p}$ & 0.2 & -0.1 & 3.4 & 6.2 & 3.4 & 6.2 & 95.9 & 96.0\tabularnewline
$2000$ & \multirow{2}{*}{Model for $H(\psi^{*})$ $(\checked)$} & $\widehat{\psi}_{\cont,1}$ & 0.1 & 0.1 & 3.3 & 5.8 & 3.3 & 5.8 & 95.2 & 95.4\tabularnewline
 &  & $\widehat{\psi}_{\cont,2}$ & 0.1 & 0.1 & 3.2 & 5.6 & 3.2 & 5.6 & 95.1 & 95.6\tabularnewline
 & \textendash{} & $\widehat{\psi}_{\disc,g}$ & 27.8 & 37.1 & 3.9 & 6.7 & 28.1 & 37.7 & 0.0 & 0.0\tabularnewline
\hline 
\multicolumn{11}{c}{Scenario (ii) with $M_{T}$ ($\times$)}\tabularnewline
\hline 
 & Model for $H(\psi^{*})$ $(\times)$ & $\widehat{\psi}_{p}$ & 7.4 & 20.2 & 5.2 & 9.9 & 9.1 & 22.5 & 68.8 & 44.6\tabularnewline
\multirow{2}{*}{$1000$} & \multirow{2}{*}{Model for $H(\psi^{*})$ $(\checked)$} & $\widehat{\psi}_{\cont,1}$ & 0.5 & 0.5 & 5.1 & 9.1 & 5.1 & 9.1 & 95.4 & 94.0\tabularnewline
 &  & $\widehat{\psi}_{\cont,2}$ & 0.5 & 0.4 & 5.1 & 9.0 & 5.1 & 9.0 & 95.0 & 95.4\tabularnewline
 & \textendash{} & $\widehat{\psi}_{\disc,g}$ & 27.7 & 38.6 & 5.9 & 10.2 & 28.4 & 40.0 & 0.2 & 3.4\tabularnewline
 & Model for $H(\psi^{*})$ $(\times)$ & $\widehat{\psi}_{p}$ & 7.4 & 20.1 & 3.5 & 6.4 & 8.1 & 21.1 & 46.2 & 17.2\tabularnewline
$2000$ & \multirow{2}{*}{Model for $H(\psi^{*})$ $(\checked)$} & $\widehat{\psi}_{\cont,1}$ & 0.4 & 0.3 & 3.4 & 5.9 & 3.4 & 5.9 & 95.0 & 95.4\tabularnewline
 &  & $\widehat{\psi}_{\cont,2}$ & 0.3 & 0.3 & 3.4 & 5.8 & 3.4 & 5.8 & 95.3 & 95.6\tabularnewline
 & \textendash{} & $\widehat{\psi}_{\disc,g}$ & 27.3 & 39.5 & 3.9 & 6.7 & 27.6 & 40.0 & 0.0 & 0.0\tabularnewline
\hline 
\end{tabular}$\checked$ (is correctly specified), $\times$ (is misspecified)
\end{table}

We consider the following estimators with details for the nuisance
models and their estimation presented in the supplementary material:
\begin{description}
\item [{(a)}] A preliminary estimator $\widehat{\psi}_{p}$ solves the
estimating equation (\ref{eq:eq4}) with $E\{H(\psi^{*})\mid\overline{V}_{u},T\geq u\}\equiv0$
and $c(\overline{V}_{u})=(1,u)^{\T}(\tau-u)I(u\leq\tau)-\E\{(1,T)^{\T}(\tau-T)I(T\leq\tau)\mid\overline{V}_{u},T\geq u\}.$
Therefore, $\widehat{\psi}_{p}$ corresponds to the proposed estimator
with a misspecified model for $E\{H(\psi^{*})\mid\overline{V}_{u},T\geq u\}$.
\item [{(b)}] The proposed estimator $\widehat{\psi}_{\cont,1}$ solves
the estimating equation (\ref{eq:eq4}), where we replace $\var\{H(\psi)\mid\overline{V}_{u},T\geq u\}$
by a constant.
\item [{(c)}] The proposed estimator $\widehat{\psi}_{\cont,2}$ solves
the estimating equation (\ref{eq:eq4}), where we obtain $\widehat{\var}\{H(\psi^{*})\mid\overline{V}_{u},T\geq u\}$
by the empirical variance of $H(\widehat{\psi}_{p})-\E\{H(\widehat{\psi}_{p})\mid\overline{V}_{u},T\geq u;\widehat{\beta}\}$,
restricted to subjects with $T\geq u$.
\item [{(d)}] The g-estimator $\widehat{\psi}_{\disc,g}$ in Section \ref{sec:discrete-SNMM}
applies to the monthly data after discretization with $24$ equally-spaced
time points from $0$ to $\tau$. For $m\geq1$, at the $m$th time
point $t_{m}$, $L_{m}$ is the the average of $L_{t}$ from $t_{m-1}\leq t\leq t_{m}$,
$A_{m}$ is the indicator of whether the treatment is initiated before
$t_{m}$, and the time to treatment initiation $T$ is $t_{m}$ if
$A_{m}=1$ and $\overline{A}_{m-1}=\overline{0}$. The g-estimator
solves the estimating equation based on (\ref{eq:semipar ee 1}),
where the nuisance models are estimated similar to what are used for
$\widehat{\psi}_{\cont,1}$ but with the re-shaped data.
\end{description}
To investigate the double robustness in Theorem \ref{Thm:2-dr}, we
consider two models for estimating $M_{T}$: the correctly specified
proportional hazards model with both time-independent and time-dependent
covariates; and the misspecified proportional hazards model with only
time-independent covariate. For all estimators, we use the bootstrap
for variance estimation with the bootstrap size $100$.

Table \ref{tab:results1} shows the simulation results in Setting
I. Under Scenario (i) when the model for the treatment process is
correctly specified, $\widehat{\psi}_{p}$, $\widehat{\psi}_{\cont,1}$
and $\widehat{\psi}_{\cont,2}$ show small biases. As a result, the
coverage rates are close to the nominal level. Under Scenario (ii)
when the model for the treatment process is misspecified, $\widehat{\psi}_{p}$
shows large biases, but $\widehat{\psi}_{\cont,1}$ and $\widehat{\psi}_{\cont,2}$
still show small biases. Moreover, the root mean squared errors of
$\widehat{\psi}_{\cont,1}$ and $\widehat{\psi}_{\cont,2}$ decrease
as the sample size increases. This confirms the double robustness
of the proposed estimators. The proposed estimator $\widehat{\psi}_{\cont,2}$
with $\widehat{\var}\{H(\psi^{*})\mid\overline{V}_{u},T\geq u\}$
produces slightly smaller standard errors; however, this reduction
is not large. In practice, we recommend $\widehat{\psi}_{\cont,1}$
because of its simpler implementation than $\widehat{\psi}_{\cont,2}$.
We note large biases in the g-estimator, which illustrates the consequence
of data pre-processing for the subsequent analysis.

\begin{table}
\caption{\label{tab:results2}Simulation results in Setting II with censoring
based on $1,000$ simulated datasets: the Monte Carlo bias, standard
error, root mean square error of the estimators, and coverage rate
of $95\%$ confidence intervals.}

\centering{}%
\begin{tabular}{cclcccccccc}
\hline 
 &  &  & \multicolumn{2}{c}{Bias ($\times10^{2}$)} & \multicolumn{2}{c}{SE ($\times10^{2}$)} & \multicolumn{2}{c}{rMSE ($\times10^{2}$)} & \multicolumn{2}{c}{CR ($\times10^{2}$)}\tabularnewline
$n$ &  & Method & $\psi_{1}^{*}$ & $\psi_{2}^{*}$ & $\psi_{1}^{*}$ & $\psi_{2}^{*}$ & $\psi_{1}^{*}$ & $\psi_{2}^{*}$ & $\psi_{1}^{*}$ & $\psi_{2}^{*}$\tabularnewline
\hline 
\multicolumn{11}{c}{Scenario (i) with $M_{T}$ ($\checked$) and $K_{C}$ $(\checked)$}\tabularnewline
\hline 
 & Model for $H(\psi^{*})$ $(\times)$ & $\widehat{\psi}_{p}$ & -0.1 & 0.2 & 5.8 & 10.9 & 5.8 & 10.9 & 95.2 & 94.8\tabularnewline
\multirow{2}{*}{$1000$} & \multirow{2}{*}{Model for $H(\psi^{*})$ $(\checked)$} & $\widehat{\psi}_{\cont,1}$ & -0.1 & 0.5 & 5.7 & 10.3 & 5.7 & 10.3 & 95.4 & 95.4\tabularnewline
 &  & $\widehat{\psi}_{\cont,2}$ & -0.1 & 0.5 & 5.6 & 10.2 & 5.6 & 10.2 & 94.5 & 95.5\tabularnewline
 & \textendash{} & $\widehat{\psi}_{\disc,g}$ & 27.7 & 32.5 & 6.7 & 12.0 & 28.5 & 34.7 & 2.4 & 24.6\tabularnewline
 & Model for $H(\psi^{*})$ $(\times)$ & $\widehat{\psi}_{p}$ & -0.3 & 0.3 & 4.2 & 7.9 & 4.2 & 7.9 & 94.6 & 94.8\tabularnewline
$2000$ & \multirow{2}{*}{Model for $H(\psi^{*})$ $(\checked)$} & $\widehat{\psi}_{\cont,1}$ & -0.3 & 0.4 & 4.2 & 7.5 & 4.2 & 7.5 & 95.0 & 94.8\tabularnewline
 &  & $\widehat{\psi}_{\cont,2}$ & -0.3 & 0.4 & 4.2 & 7.4 & 4.2 & 7.4 & 95.1 & 95.0\tabularnewline
 & \textendash{} & $\widehat{\psi}_{\disc,g}$ & 27.5 & 32.9 & 4.7 & 8.2 & 27.9 & 33.9 & 0.0 & 1.6\tabularnewline
\hline 
\multicolumn{11}{c}{Scenario (ii) with $M_{T}$ $(\times)$ and $K_{C}$ $(\checked)$}\tabularnewline
\hline 
 & Model for $H(\psi^{*})$ $(\times)$ & $\widehat{\psi}_{p}$ & 7.0 & 21.0 & 6.0 & 11.4 & 9.2 & 23.9 & 82.2 & 57.8\tabularnewline
\multirow{2}{*}{$1000$} & \multirow{2}{*}{Model for $H(\psi^{*})$ $(\checked)$} & $\widehat{\psi}_{\cont,1}$ & -0.1 & 1.1 & 5.7 & 10.3 & 5.7 & 10.4 & 95.0 & 95.2\tabularnewline
 &  & $\widehat{\psi}_{\cont,2}$ & -0.1 & 1.1 & 5.5 & 10.1 & 5.5 & 10.2 & 95.2 & 95.3\tabularnewline
 & \textendash{} & $\widehat{\psi}_{\disc,g}$ & 27.4 & 33.4 & 6.7 & 12.0 & 28.2 & 35.5 & 3.2 & 22.2\tabularnewline
 & Model for $H(\psi^{*})$ $(\times)$ & $\widehat{\psi}_{p}$ & 7.0 & 21.2 & 4.2 & 8.2 & 8.2 & 22.8 & 63.6 & 29.2\tabularnewline
$2000$ & \multirow{2}{*}{Model for $H(\psi^{*})$ $(\checked)$} & $\widehat{\psi}_{\cont,1}$ & -0.3 & 1.0 & 4.1 & 7.5 & 4.2 & 7.6 & 94.4 & 95.2\tabularnewline
 &  & $\widehat{\psi}_{\cont,2}$ & -0.3 & 1.1 & 4.0 & 7.4 & 4.1 & 7.5 & 94.7 & 95.4\tabularnewline
 & \textendash{} & $\widehat{\psi}_{\disc,g}$ & 27.2 & 33.7 & 4.7 & 8.1 & 27.6 & 34.7 & 0.0 & 1.2\tabularnewline
\hline 
\multicolumn{11}{c}{Scenario (iii) with $M_{T}$ $(\checked)$ and $K_{C}$ $(\times)$}\tabularnewline
\hline 
 & Model for $H(\psi^{*})$ $(\times)$ & $\widehat{\psi}_{p}$ & -0.1 & 0.2 & 5.8 & 11.0 & 5.8 & 11.0 & 95.0 & 95.0\tabularnewline
\multirow{2}{*}{$1000$} & \multirow{2}{*}{Model for $H(\psi^{*})$ $(\checked)$} & $\widehat{\psi}_{\cont,1}$ & -0.1 & 0.4 & 5.7 & 10.4 & 5.7 & 10.4 & 95.2 & 95.6\tabularnewline
 &  & $\widehat{\psi}_{\cont,2}$ & -0.1 & 0.3 & 5.7 & 10.4 & 5.7 & 10.4 & 95.0 & 95.3\tabularnewline
 & \textendash{} & $\widehat{\psi}_{\disc,g}$ & 27.7 & 32.3 & 6.7 & 12.1 & 28.5 & 34.5 & 1.8 & 26.2\tabularnewline
 & Model for $H(\psi^{*})$ $(\times)$ & $\widehat{\psi}_{p}$ & -0.3 & 0.4 & 4.2 & 7.9 & 4.3 & 7.9 & 95.0 & 94.8\tabularnewline
$2000$ & \multirow{2}{*}{Model for $H(\psi^{*})$ $(\checked)$} & $\widehat{\psi}_{\cont,1}$ & -0.3 & 0.4 & 4.2 & 7.5 & 4.2 & 7.6 & 95.2 & 95.4\tabularnewline
 &  & $\widehat{\psi}_{\cont,2}$ & -0.3 & 0.4 & 4.1 & 7.2 & 4.1 & 7.2 & 95.4 & 95.2\tabularnewline
 & \textendash{} & $\widehat{\psi}_{\disc,g}$ & 27.4 & 32.6 & 4.7 & 8.2 & 27.8 & 33.7 & 0.0 & 1.8\tabularnewline
\hline 
\multicolumn{11}{c}{Scenario (iv) with $M_{T}$ ($\times$) and $K_{C}$ $(\times)$}\tabularnewline
\hline 
 & Model for $H(\psi^{*})$ $(\times)$ & $\widehat{\psi}_{p}$ & 6.9 & 20.5 & 5.9 & 11.3 & 9.1 & 23.5 & 81.0 & 58.6\tabularnewline
\multirow{2}{*}{$1000$} & \multirow{2}{*}{Model for $H(\psi^{*})$ $(\checked)$} & $\widehat{\psi}_{\cont,1}$ & -0.0 & 1.0 & 5.7 & 10.4 & 5.7 & 10.4 & 94.8 & 95.0\tabularnewline
 &  & $\widehat{\psi}_{\cont,2}$ & -0.0 & 1.0 & 5.5 & 10.3 & 5.5 & 10.3 & 95.0 & 95.2\tabularnewline
 & \textendash{} & $\widehat{\psi}_{\disc,g}$ & 27.5 & 33.1 & 6.8 & 12.1 & 28.3 & 35.2 & 3.0 & 24.0\tabularnewline
 & Model for $H(\psi^{*})$ $(\times)$ & $\widehat{\psi}_{p}$ & 6.9 & 20.8 & 4.1 & 8.1 & 8.1 & 22.3 & 63.4 & 30.2\tabularnewline
$2000$ & \multirow{2}{*}{Model for $H(\psi^{*})$ $(\checked)$} & $\widehat{\psi}_{\cont,1}$ & -0.2 & 0.9 & 4.2 & 7.5 & 4.2 & 7.6 & 94.2 & 95.4\tabularnewline
 &  & $\widehat{\psi}_{\cont,2}$ & -0.2 & 0.8 & 4.1 & 7.4 & 4.1 & 7.4 & 94.6 & 95.6\tabularnewline
 & \textendash{} & $\widehat{\psi}_{\disc,g}$ & 27.2 & 33.4 & 4.7 & 8.1 & 27.6 & 34.4 & 0.0 & 1.6\tabularnewline
\hline 
\end{tabular}

$\checked$ (is correctly specified), $\times$ (is misspecified)
\end{table}
In Setting II, we further generate the time to censoring $C$ with
the hazard rate $\lambda_{C}(t\mid\overline{V}_{t})=\lambda_{C,0}(t)\exp(\eta_{1}L_{TI}+\eta_{2}L_{TD,t})$,
with $\lambda_{C,0}(t)=0.2$, and $\eta_{1}=\eta_{2}=0.2$. In the
presence of censoring, we consider the four estimators (a)\textendash (d)
considered in Setting I with weighting; i.e., the corresponding estimating
functions are now weighted by $\delta_{C}/\widehat{K}_{C}(\tau\mid\overline{V}_{\tau})$.
To investigate the multiple robustness in Theorem \ref{Thm:3-mr},
we additionally consider two models for estimating $K_{C}$: the correctly
specified proportional hazards model with both time-independent and
time-dependent covariates; and the misspecified proportional hazards
model without covariate.

Table \ref{tab:results2} shows the simulation results in Setting
II. Under Scenarios (i) and (iii) when the model for the treatment
process is correctly specified, $\widehat{\psi}_{p}$, $\widehat{\psi}_{\cont,1}$
and $\widehat{\psi}_{\cont,2}$ show small biases, regardless whether
the models for $H(\psi^{*})$ and the censoring process are correctly
specified or not. Moreover, under Scenarios (ii) and (iv) when the
model for the treatment process is misspecified, $\widehat{\psi}_{p}$
shows large biases, but as predicted by the multiple robustness, $\widehat{\psi}_{\cont,1}$
and $\widehat{\psi}_{\cont,2}$ still show small biases. Again, the
discretized g-estimator shows large biases across all scenarios.

\section{Estimating the effect of time to initiating HAART \label{sec:Application}}

\subsection{Acute infection and early disease research program}

\begin{figure}
\caption{\label{fig:irregular visit}CD4 count and log viral load for $5$
random patients measured at irregularly spaced time points, which
are colored by patients.}

\centering{}\includegraphics[scale=0.24]{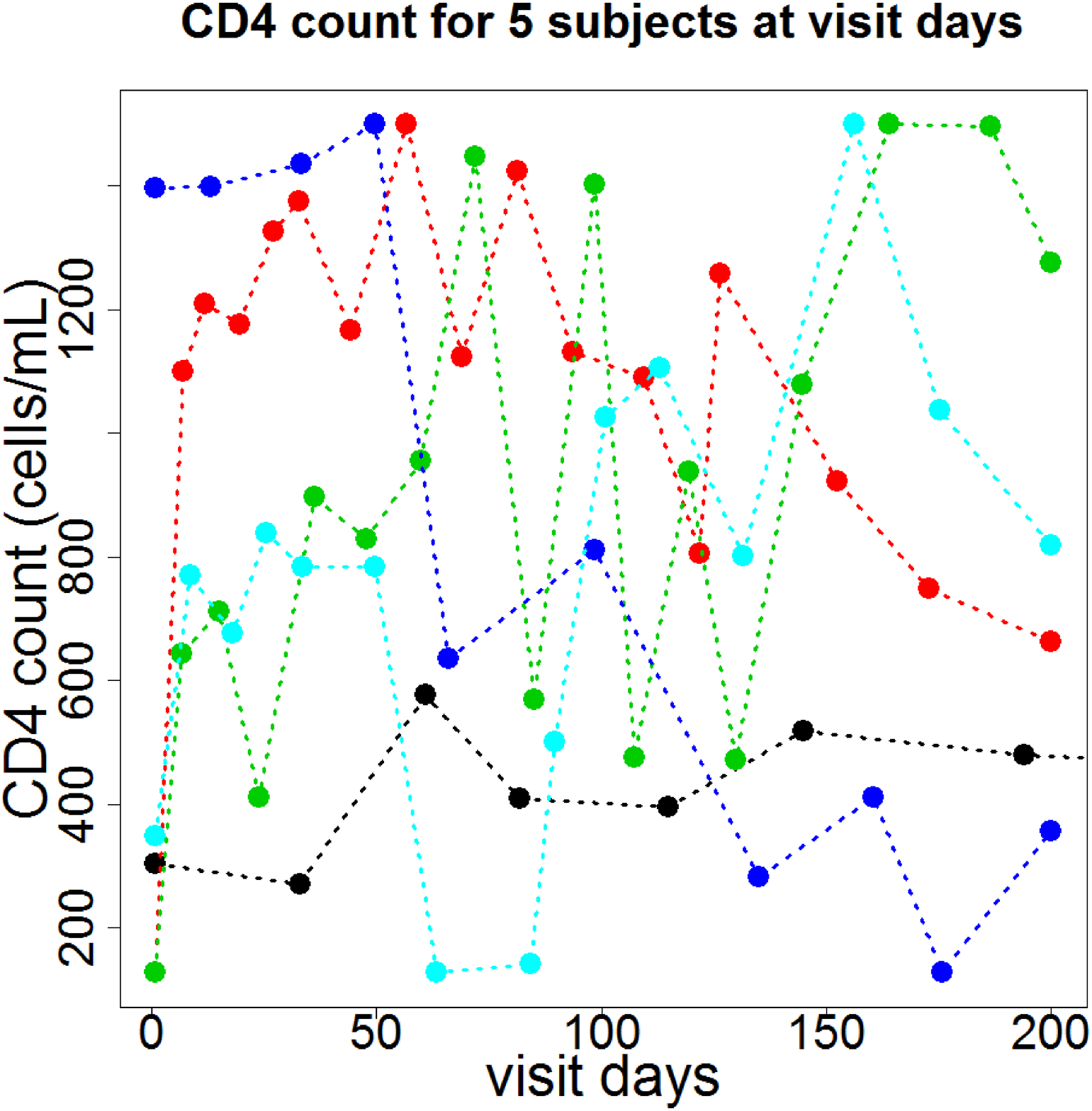}\includegraphics[scale=0.24]{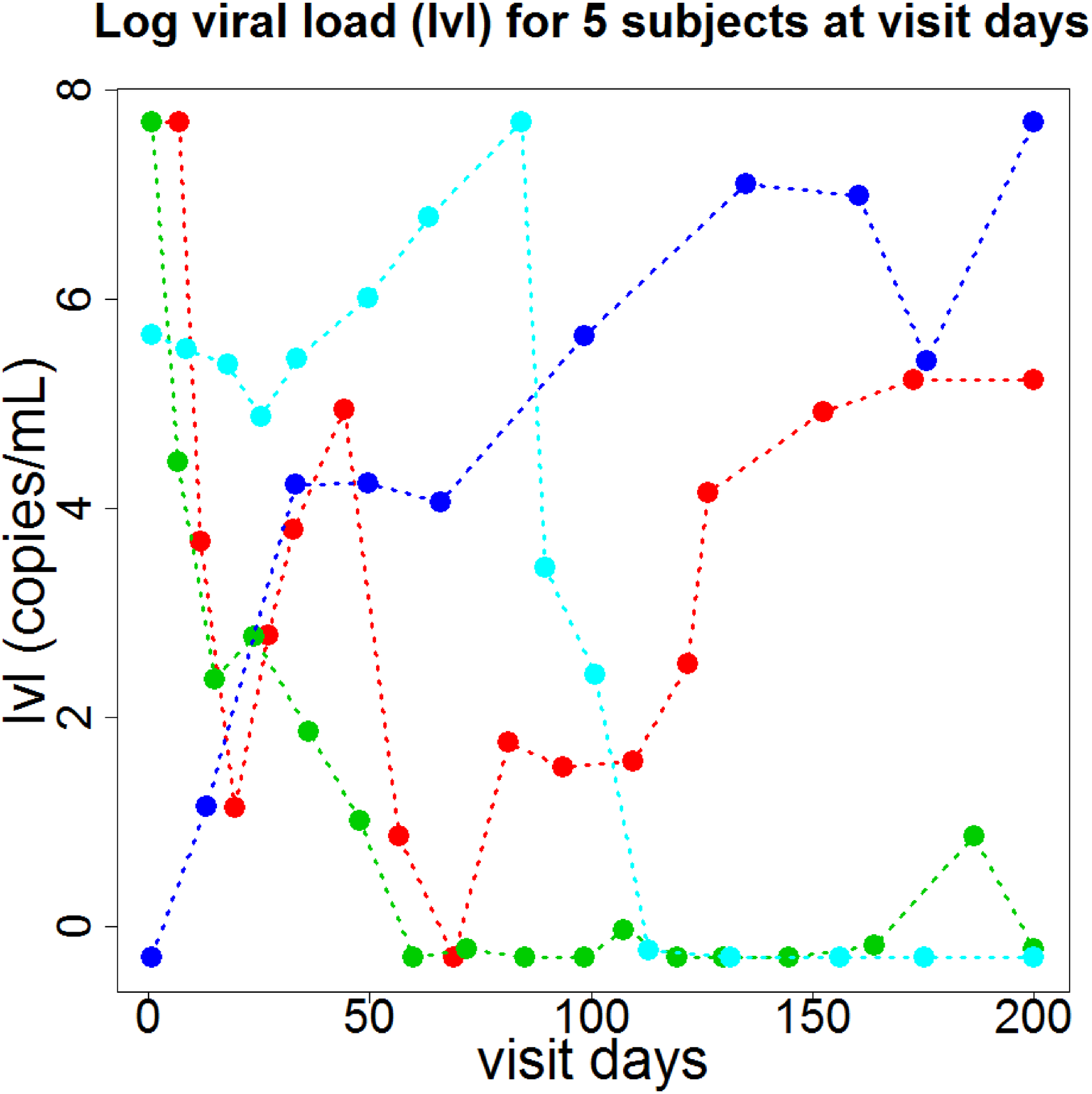}
\end{figure}

We apply our method to the observational AIEDRP database consisting
of $1762$ HIV-positive patients diagnosed during acute and early
infection. \citet{lok2012impact} investigated how the time to initiation
of HAART after HIV infection predicts the effect of one year of treatment
based on this database. \citet{yang2015gof,yang2017sensitivity} developed
a goodness-of-fit procedure to assess the treatment effect model and
a sensitivity analysis to the departure of the NUC assumption. All
these methods were based on the monthly data after discretization.
However, the observations from the original data are collected by
user-initiated visits and are irregularly spaced \citep{hecht2006multicenter}.
Figure \ref{fig:irregular visit} shows the visit times for $5$ random
patients. As can be seen, we have irregular visits, and the number
and frequency of visits vary from patients to patients.

\subsection{Objective}

We aim to estimate the averaged causal effect of the time to HAART
initiation on the mean CD4 count at year 2 after HIV infection directly
on the basis of the original data without discretization. We assume
a continuous-time SNMM $\gamma(\overline{V}_{u};\psi^{*})=(\psi_{1}^{*}+\psi_{2}^{*}t)(\tau-t)I(t\leq\tau)$.
As discussed before, $\psi_{2}^{*}$ quantifies the impact of time
to treatment initiation. The rationale for this modeling choice is
because the duration of treatment may well be predictive of its effect.

\subsection{Estimator and nuisance models}

We consider the proposed estimators $\widehat{\psi}_{\cont,1}$ and
$\widehat{\psi}_{\cont,2}$ specified in Section \ref{sec:simulation}.
The estimation procedure requires specifying and fitting nuisance
models, which we now consider. 

\textit{Model for the treatment process.} The model for the treatment
process $(M_{T})$ is a time-dependent proportional hazards model
adjusting for gender, age (age at infection), race (white non-Hispanic
race), injdrug (injection drug ever/never), CD4$_{u}^{1/2}$ (square
root of current CD4 count ), lvl$_{u}$ (log viral load), days from
last visit$_{u}$ (number of days since the last visit), first visit$_{u}$
(whether the visit is the first visit), second visit$_{u}$ (whether
the visit is the second visit). Table \ref{tab:AIEDRP2} (the left
portion) reports the point and standard error estimates of coefficients
in the treatment process model. Male and injection drug user are negatively
associated with the hazard of treatment initiation, which are significant
at the $0.05$ level. Moreover, higher CD4 count and viral load, more
days from the last visit, and whether the visit is the first visit
are associated with a decreased hazard of treatment initiation.

\textit{Model for the censoring process.} The model for the censoring
process $(K_{C})$ is a time-dependent proportional hazards model
adjusting for gender, age, white non-Hispanic race, injdrug, CD4$_{u}^{1/2}$,
lvl$_{u}$, and Treated$_{u}$ (whether a patient had initiated HAART).
Table \ref{tab:AIEDRP2} (the right portion) reports the point and
standard error estimates of coefficients in the censoring model. Age
is negatively associated with the hazard of censoring, while being
an injection drug user is positively associated with the hazard of
censoring, both are highly significant. Moreover, higher CD4 count,
more days from the last visit, and whether the visit is the first
visit are highly associated with a decreased hazard of censoring.

\begin{table}
\caption{\label{tab:AIEDRP2}Fitted time-dependent proportional hazards models
for time to treatment initiation and time to censoring}

\centering{}%
\begin{tabular}{ccccccccc}
\hline 
 & \multicolumn{4}{c}{time to treatment initiation} & \multicolumn{3}{c}{time to censoring} & \tabularnewline
\hline 
 & Est & SE & p-val &  & Est & SE & p-val & \tabularnewline
\hline 
male & -0.35 & 0.161 & 0.03 & {*} & 0.21 & 0.159 & 0.19 & \tabularnewline
age & 0.01 & 0.003 & 0.08 & . & -0.02 & 0.004 & 0.00 & {*}{*}{*}\tabularnewline
white non-hispanic & 0.12 & 0.066 & 0.07 & . & 0.02 & 0.077 & 0.77 & \tabularnewline
injdrug & -0.50 & 0.180 & 0.01 & {*}{*} & 0.74 & 0.156 & 0.00 & {*}{*}{*}\tabularnewline
CD4$_{u}^{1/2}$ & -0.06 & 0.007 & 0.00 & {*}{*}{*} & -0.03 & 0.007 & 0.00 & {*}{*}{*}\tabularnewline
lvl$_{u}$ & -0.14 & 0.013 & 0.00 & {*}{*}{*} & 0.04 & 0.016 & 0.02 & {*}\tabularnewline
days from last visit$_{u}$ & -0.03 & 0.002 & 0.00 & {*}{*}{*} & -0.01 & 0.001 & 0.00 & {*}{*}{*}\tabularnewline
first visit$_{u}$ & -3.06 & 0.111 & 0.00 & {*}{*}{*} & -1.24 & 0.231 & 0.00 & {*}{*}{*}\tabularnewline
second visit$_{u}$ & -0.04 & 0.081 & 0.61 &  & 0.68 & 0.178 & 0.00 & {*}{*}{*}\tabularnewline
Treated$_{u}$ & \textendash{} & \textendash{} & \textendash{} &  & -0.15 & 0.102 & 0.15 & \tabularnewline
\hline 
\end{tabular}

Signif. codes: 0 '{*}{*}{*}' 0.001 '{*}{*}' 0.01 '{*}' 0.05 '.'
\end{table}

\textit{Model for the potential outcome mean function.} The outcome
model $E\{H(\widehat{\psi}_{p})\mid\overline{V}_{u},T\geq u;\beta\}$
is a linear regression model where the covariates include age, male,
race, injdrug, CD4$_{u}$, lvl$_{u}$, CD4$_{u}^{3/4}(\tau-u)$, CD4$_{u}^{3/4}\times(\tau-u)\times$age,
CD4$_{u}^{3/4}\times(\tau-u)\times$male, CD4$_{u}^{3/4}\times(\tau-u)\times$race,
CD4$_{u}^{3/4}\times(\tau-u)\times$injdrug, CD4$_{u}^{3/4}\times(\tau-u)\times$lvl$_{u}$,
CD4slope$_{u}$ measured, CD4slope$_{u}\times(\tau-u)^{1/2}$ $I(u\leq6)(6-u)$,
and $I(u\leq6)(36-u^{2})$. This model specification is motivation
based on the substantive literature including \citet{taylor1994stochastic,taylor1998does,rodriguez2006predictive,may2009cd4}.\textbf{}

\textit{Other nuisance models}. $E(\tau-T\mid\overline{L}_{u},T\geq u)$
and $E\{T(\tau-T)\mid\overline{L}_{u},T\geq u)\}$ are linear regression
models where the covariates include $u$, $(\tau-u)$, male$\times(\tau-u)$,
age$\times(\tau-u)$, race$\times(\tau-u)$, injdrug$\times(\tau-u)$,
CD4$_{u}^{1/2}$$\times(\tau-u)$, lvl$_{u}$$\times(\tau-u)$, days
from last visit$_{u}\times(\tau-u)$, first visit$_{u}\times(\tau-u)$,
second visit$_{u}\times(\tau-u)$.

The confounding variables and nuisance models are chosen on the basis
of the substantive knowledge and the established literature, and therefore
the NUC assumption is plausible in this application. We use bootstrap
for variance estimation with the bootstrap size $100$ and compute
the $95\%$ Wald confidence interval.

\subsection{Results}

Table \ref{tab:Results-2} shows the results for the effect of time
to HAART initiation on the CD4 count at year 2. We note only slight
differences in the point estimates between our estimators. Based on
our results, on average, initiation of HAART at the time of infection
($t=0$) can increase CD4 counts at year 2 by $14.1\text{cells/mm\ensuremath{^{3}} per month}\times24\text{ months}\approx338$
cells/mm$^{3}$; while initiation of HAART $3$ months after the time
of infection can increase CD4 counts at year 2 by $(14.1-1.00\times3)\times(24-3)\approx233$
cells/mm$^{3}$.

\begin{table}
\caption{\label{tab:Results-2}Results of the effect of time to HAART initiation
on the CD4 count at year 2}

\centering{}%
\begin{tabular}{lccccc}
\hline 
Method & Est  & SE  & lower .95  & upper .95 & p-val \tabularnewline
\hline 
 & \multicolumn{5}{c}{$\psi_{1}^{*}$ cells/mm$^{3}$ per month}\tabularnewline
Proposed 1: $\widehat{\psi}_{\cont,1}$  & 14.1  & 1.1  & 12.0  & 16.3  & 0.000 \tabularnewline
Proposed 2: $\widehat{\psi}_{\cont,2}$  & 14.3  & 1.1  & 12.2  & 16.6  & 0.000 \tabularnewline
\hline 
 & \multicolumn{5}{c}{$\psi_{2}^{*}$ cells/mm$^{3}$ per month$^{2}$}\tabularnewline
Proposed 1: $\widehat{\psi}_{\cont,1}$  & -1.00  & 0.23  & -1.42  & -0.50  & 0.000\tabularnewline
Proposed 2: $\widehat{\psi}_{\cont,2}$  & -1.01  & 0.23  & -1.43  & -0.52  & 0.000\tabularnewline
\hline 
\end{tabular}
\end{table}

\section{Discussion \label{sec:Discussion}}

In this article, we have developed a new semiparametric estimation
framework for continuous-time SNMMs to evaluate treatment effects
with irregularly spaced longitudinal observations. Our approach does
not require specifying the joint distribution of the covariate, treatment,
outcome and censoring processes. Moreover, our method achieves a multiple
robustness property requiring the correct specification of either
the model for the potential outcome mean function or the model for
the treatment process, regardless whether the censoring process model
is correctly specified. This robustness property will be useful when
there is little prior or substantive knowledge about the data processes.
Below, we discuss several directions for future work.

\subsection{Other types of outcome}

To accommodate different types of outcome, we consider a general specification
of the continuous-time SNMM as
\begin{equation}
\gamma_{t}(\overline{L}_{t})=g\left[\E\left\{ Y^{(t)}\mid\overline{L}_{t},T\geq t\right\} \right]-g\left[\E\left\{ Y^{(\infty)}\mid\overline{L}_{t},T\geq t\right\} \right]=\gamma_{t}(\overline{L}_{t};\psi^{*}),\label{eq:g-SNMM-1}
\end{equation}
where $g(\cdot)$ is a pre-specified link function. For the continuous
outcome, $g(\cdot)$ can be an identity link, i.e. $g(x)=x$, as we
adopt in this article. For the binary outcome, $g(\cdot)$ can be
a logit link, i.e. $g(x)=\logit(x)\coloneqq\log\{x/(1-x)\}$. Then,
(\ref{eq:g-SNMM-1}) specifies the treatment effect on the odds ratio
scale, i.e. $\text{odds}\left\{ Y^{(t)}\mid\overline{L}_{t},T\geq t\right\} /\text{odds}\left\{ Y^{(\infty)}\mid\overline{L}_{t},T\geq t\right\} $,
where $\text{odds}(Y\mid X)=P(Y=1\mid X)/P(Y=0\mid X)$. In this case,
$H(\psi^{*})$ can be constructed as $H(\psi^{*})=\text{expit}\left[\logit\{\E(Y\mid\overline{L}_{t},T\geq t)\}-\gamma_{t}(\overline{L}_{t};\psi^{*})\right]$.
We can develop the corresponding semiparametric efficiency theory
for $\psi^{*}$ similarly. For a time to event outcome, we can consider
the structural nested failure time models \citep{robins1991correcting,robins1992estimation,yang2018semiparametric}.

\subsection{Effect modification and model selection\label{subsec:Effect-modification}}

Effect modification occurs when the magnitude of the treatment effect
varies as a function of observed covariates. To allow for time-varying
treatment effect modifiers, assume $\gamma_{t}(\overline{V}_{t};\psi^{*})=\{\psi_{1}^{*}+\psi_{2}^{*}t+\psi_{3}^{*\T}W(t,\overline{V}_{t})\}(\tau-t)I(t\leq\tau)$,
when $W(t,\overline{V}_{t})$ is a pre-specified and possibly high-dimensional
function of $t$ and $\overline{V}_{t}$. It is important to identify
the true treatment effect modifiers, which can facilitate development
of optimal treatment strategies in personalized medicine \citep{murphy2003optimal}.
We will develop a variable selection procedure for identifying effect
modifiers. The insight is that we have a larger number of estimating
functions than the number of parameters. The problem for effect modifiers
selection falls into the recent work of \citet{chang2017new} on high-dimensional
statistical inferences with over-identification.

\subsection{Sensitivity analysis to the NUC assumption}

The key assumption to identify the causal parameters in the continuous-time
SNMM is the NUC assumption. However, this assumption is not verifiable
based on the observed data. In future studies, it is desirable that
the follow-up visits and treatment assignment be determined by study
protocol. By formalizing the visit process and treatment assignment,
one knows by design which covariates contribute to the treatment process
to ensure the NUC assumption holds with all the relevant covariates.
In the absence of study protocol, we then recommend conducting sensitivity
analysis to assess the impact of possible uncontrolled confounding.
For the discrete-time SNMMs, \citet{yang2017sensitivity} assumed
a bias function $b(\overline{L}_{m})=\E\{Y^{(\infty)}\mid\overline{A}_{m-1}=\overline{0},A_{m}=1,\overline{L}_{m}\}-\E\{Y^{(\infty)}\mid\overline{A}_{m-1}=\overline{0},A_{m}=0,\overline{L}_{m}\}$
that quantifies the impact of unmeasured confounding and developed
a modified g-estimator. For the continuous-time SNMMs, it would also
be important to develop a sensitivity analysis methodology, along
the lines of \citet{robins1999sensitivity} or \citet{yang2017sensitivity},
to evaluate the sensitivity of causal inference to departures from
the NUC assumption.

\section*{Acknowledgment}

The author would like to thank Anastasio A. Tsiatis for insightful
and fruitful discussions. Dr. Yang is partially supported by NSF DMS
1811245 and NCI P01 CA142538.

\section*{Supplementary Material}

Supplementary material online includes proofs, technical and simulation
details. 

\bibliographystyle{dcu}
\bibliography{ci_contiSNMMjasa}

\newpage{}

\appendix

\global\long\def\theequation{S\arabic{equation}}%
 \setcounter{equation}{0}

\global\long\def\thelemma{S\arabic{lemma}}%
 \setcounter{lemma}{0}

\global\long\def\theexample{S\arabic{example}}%
 \setcounter{equation}{0}

\global\long\def\thesection{S\arabic{section}}%
 \setcounter{section}{0}

\global\long\def\thetheorem{S\arabic{theorem}}%
 \setcounter{equation}{0}

\global\long\def\thecondition{S\arabic{condition}}%
 \setcounter{equation}{0}

\global\long\def\theremark{S\arabic{remark}}%
 \setcounter{equation}{0}

\global\long\def\thestep{S\arabic{step}}%
 \setcounter{equation}{0}

\global\long\def\theassumption{S\arabic{assumption}}%
 \setcounter{assumption}{0}

\global\long\def\theproof{S\arabic{proof}}%
 \setcounter{equation}{0}

\global\long\def\theproposition{S{proposition}}%
 \setcounter{equation}{0}

\textbf{\large{}Supplementary material for ``Structural nested mean
models with irregularly spaced observations''}{\large\par}

\section{Proofs}

\subsection{Proof of (\ref{eq:UNC2})}

First, we express
\begin{eqnarray*}
 &  & \lambda_{T}\{t\mid\overline{V}_{t},Y^{(\infty)}\}=\lim_{h\rightarrow0}h^{-1}P\{t\leq T<t+h\mid\overline{V}_{t},Y^{(\infty)},T\geq t\}\\
 & = & \lim_{h\rightarrow0}h^{-1}\frac{f\{Y^{(\infty)}\mid\overline{V}_{t},t\leq T<t+h\}P\{t\leq T<t+h\mid\overline{V}_{t},T\geq t\}}{f\{Y^{(\infty)}\mid\overline{V}_{t},T\geq t\}}\\
 & = & \lim_{h\rightarrow0}h^{-1}\frac{f\{H(\psi^{*})\mid\overline{V}_{t},t\leq T<t+h\}P\{t\leq T<t+h\mid\overline{V}_{t},T\geq t\}}{f\{H(\psi^{*})\mid\overline{V}_{t},T\geq t\}}\\
 & = & \lim_{h\rightarrow0}h^{-1}P\{t\leq T<t+h\mid\overline{V}_{t},H(\psi^{*}),T\geq t\}\\
 & = & \lambda_{T}\{t\mid\overline{V}_{t},H(\psi^{*})\},
\end{eqnarray*}
where the second equality follows by the Bayes rule, and the third
equality follows by Model (\ref{eq:cont-SNMM}) which implies that
the distribution of $\{\overline{V}_{t},Y^{(\infty)}\}$ is the same
as the distribution of $\{\overline{V}_{t},H(\psi^{*})\}$.

Second, by Assumption \ref{asumption:CT-UNC}, $\lambda_{T}\{t\mid\overline{V}_{t},Y^{(\infty)}\}=\lambda_{T}(t\mid\overline{V}_{t})$.
Therefore, $\lambda_{T}\{t\mid\overline{V}_{t},H(\psi^{*})\}=\lambda_{T}\{t\mid\overline{V}_{t},Y^{(\infty)}\}=\lambda_{T}(t\mid\overline{V}_{t})$.

\subsection{Proof of Theorem \ref{Thm: cont-nuisance}}

First, we characterize the semiparametric likelihood function of
$\psi^{*}$ based on a single variable $O=(\overline{V}_{\tau},Y)$.
The semiparametric likelihood is 
\begin{equation}
f_{O}\left(\overline{V}_{\tau},Y\right)=\left\{ \frac{\de H(\psi^{*})}{\de Y}\right\} f_{\{\overline{V}_{\tau},H(\psi^{*})\}}\{\overline{V}_{\tau},H(\psi^{*})\}=f_{\{\overline{V}_{\tau},H(\psi^{*})\}}\{\overline{V}_{\tau},H(\psi^{*})\},\label{eq:slik}
\end{equation}
where the first equality follows by the transformation of $O$ to
$\{\overline{V}_{\tau},H(\psi^{*})\}$, and the second equality follows
because $\de H(\psi^{*})/\de Y=1$. To express (\ref{eq:slik}) further,
we let the observed times to treatment initiation among the $n$ subjects
be $v_{0}=0<v_{1}<\cdots<v_{M}$. By Assumption \ref{asumption:CT-UNC}
and (\ref{eq:UNC2}), we express
\begin{eqnarray}
f_{O}\left(\overline{V}_{\tau},Y;\psi^{*},\theta\right) & = & f\left\{ H(\psi^{*});\theta_{1}\right\} \prod_{k=1}^{M}f\left\{ L_{v_{k}}\mid\overline{A}_{v_{k-1}}=\overline{0},\overline{L}_{v_{k-1}},H(\psi^{*});\theta_{2}\right\} \nonumber \\
 &  & \times\prod_{v=v_{1}}^{v_{M}}f\left\{ A_{v_{k}}\mid\overline{A}_{v_{k-1}}=\overline{0},\overline{L}_{v_{k}},H(\psi^{*});\theta_{3}\right\} ,\nonumber \\
 & = & f\left\{ H(\psi^{*});\theta_{1}\right\} \prod_{k=1}^{M}f\left\{ L_{v_{k}}\mid\overline{A}_{v_{k-1}}=\overline{0},\overline{L}_{v_{k-1}},H(\psi^{*});\theta_{2}\right\} \nonumber \\
 &  & \times\prod_{v=v_{1}}^{v_{M}}f\left\{ A_{v_{k}}\mid\overline{A}_{v_{k-1}}=\overline{0},\overline{L}_{v_{k}};\theta_{3}\right\} \nonumber \\
 & = & f\left\{ H(\psi^{*});\theta_{1}\right\} \prod_{k=1}^{M}f\left\{ L_{v_{k}}\mid\overline{A}_{v_{k-1}}=\overline{0},\overline{L}_{v_{k-1}},H(\psi^{*});\theta_{2}\right\} \nonumber \\
 &  & \times f(T,\Gamma\mid\overline{V}_{T};\theta_{3}),\label{eq:slik2}
\end{eqnarray}
where $\theta=(\theta_{1},\theta_{2},\theta_{3})$ is a vector of
the infinite-dimensional nuisance parameters given the nonparametric
models, and the third equality follows because $\prod_{k=1}^{M}f\left(A_{v_{k}}\mid\overline{A}_{v_{k-1}}=\overline{0},\overline{L}_{v_{k}};\theta_{3}\right)$
can be equivalently expressed as the likelihood based on the data
$(T,\Gamma)$ given $\overline{V}_{T}$. 

Second, we characterize $\Lambda_{k}$, the nuisance tangent space
for $\theta_{k},$ for $k=1,2,3$. Assuming $f\left\{ H(\psi^{*});\theta_{1}\right\} $
and $\prod_{k=1}^{M}f\left\{ L_{v_{k}}\mid\overline{A}_{v_{k-1}}=\overline{0},\overline{L}_{v_{k-1}},H(\psi^{*});\theta_{2}\right\} $
are nonparametric, it follows from Section 4.4 of \citet{tsiatis2007semiparametric}
that the tangent space regarding $\theta_{1}$ is
\[
\Lambda_{1}=\left\{ s\left\{ H(\psi^{*})\right\} \in\mathbb{\R}^{p}:\E\left[s\left\{ H(\psi^{*})\right\} \right]=0\right\} ,
\]
and the tangent space of $\theta_{2}$ is 
\begin{multline*}
\Lambda_{2}=\sum_{k=1}^{M}\left\{ S\left\{ \overline{V}_{v_{k}-1},L_{v_{k}},H(\psi^{*})\right\} \in\R^{p}:\right.\\
\left.\E\left[S\left\{ \overline{V}_{v_{k}-1},L_{v_{k}},H(\psi^{*})\right\} \mid\overline{A}_{v_{k-1}}=\overline{0},\overline{L}_{v_{k-1}},H(\psi^{*})\right]=0\right\} .
\end{multline*}
By writing
\begin{multline*}
f_{(T,\Gamma\mid\overline{V}_{T})}(T,\Gamma\mid\overline{V}_{T})=\lambda_{T}(T\mid\overline{V}_{T})^{\Gamma}\exp\left\{ -\int_{0}^{T}\lambda_{T}(u\mid\overline{V}_{u})\de u\right\} \\
\times\left\{ f_{T\mid\overline{V}_{T}}(T\mid\overline{V}_{T})\right\} ^{1-\Gamma}\left\{ \int_{T}^{\infty}f_{T\mid\overline{V}_{T}}(u\mid\overline{V}_{u})\de u\right\} ^{\Gamma},
\end{multline*}
it follows from \citet{tsiatis2007semiparametric} that the tangent
space of $\theta_{3}$ is 
\[
\Lambda_{3}=\left\{ \int h_{u}(\overline{V}_{u})\de M_{T}(u):\text{for all }h_{u}(\overline{V}_{u})\in\mathbb{\R}^{p}\right\} .
\]
Then, the nuisance tangent space becomes $\Lambda=\Lambda_{1}\oplus\Lambda_{2}\oplus\Lambda_{3}$,
where $\oplus$ denotes a direct sum. This is because $\theta_{1},$
$\theta_{2},$ and $\theta_{3}$ separate out in the likelihood function
and therefore $\Lambda_{1}$, $\Lambda_{2}$ and $\Lambda_{3}$ are
mutually orthogonal.

Third, we characterize $\Lambda^{\bot}$ using the following technical
trick. Define
\[
\Lambda_{3}^{*}=\left\{ \int h_{u}\{\overline{V}_{u},H(\psi^{*})\}\de M_{T}(u):\ h_{u}\{\overline{V}_{u},H(\psi^{*})\}\in\R^{p}\right\} .
\]
Because the tangent space $\Lambda_{1}\oplus\Lambda_{2}\oplus\Lambda_{3}^{*}$
is that for a nonparametric model; i.e., a model that allows for all
densities of $O$, and because the tangent space for a nonparametric
model is the entire Hilbert space, we obtain that $\mathcal{H}=\Lambda_{1}\oplus\Lambda_{2}\oplus\Lambda_{3}^{*}.$
Because $\Lambda^{\bot}$ must be orthogonal to $\Lambda_{1}\oplus\Lambda_{2}$,
$\Lambda^{\bot}$ consists of all elements of $\Lambda_{3}^{*}$ that
are orthogonal to $\Lambda_{3}$. It then suffices to find the projection
of all elements of $\Lambda_{3}^{*}$, $\int h_{u}\{\overline{V}_{u},H(\psi^{*})\}\de M_{T}(u)$,
onto $\Lambda_{3}^{\bot}$. To find the projection, we derive $h_{u}^{*}(\overline{V}_{u})$
such that 
\[
\left[\int h_{u}\{\overline{V}_{u},H(\psi^{*})\}\de M_{T}(u)-\int h_{u}^{*}(\overline{V}_{u})\de M_{T}(u)\right]\in\Lambda_{3}^{\bot}.
\]
Therefore, we have 
\begin{equation}
\E\left(\int\left[h_{u}\{\overline{V}_{u},H(\psi^{*})\}-h_{u}^{*}(\overline{V}_{u})\right]\de M_{T}(u)\times\int h_{u}(\overline{V}_{u})\de M_{T}(u)\right)=0,\label{eq:eq1}
\end{equation}
for any $h_{u}(\overline{V}_{u})$. It is important to note that by
Assumption \ref{asumption:CT-UNC}, $M_{T}(t)$ is a martingale with
respect to the filtration $\sigma\{\overline{V}_{t},H(\psi^{*})\}$.
If $P_{1}(u)$ and $P_{2}(u)$ are locally bounded $\sigma\{\overline{V}_{t},H(\psi^{*})\}$-predictable
processes, then we have the following useful result:
\begin{equation}
\E\left\{ \int_{0}^{t}P_{1}(u)\de M_{T}(u)\int_{0}^{t}P_{2}(u)\de M_{T}(u)\right\} =\int_{0}^{t}P_{1}(u)P_{2}(u)\lambda_{T}(u\mid\overline{V}_{u})Y_{T}(u)\de u.\label{eq:lemma1}
\end{equation}
By (\ref{eq:lemma1}), (\ref{eq:eq1}) becomes 
\begin{multline*}
\E\left(\int\left[h_{u}\{\overline{V}_{u},H(\psi^{*})\}-h_{u}^{*}(\overline{V}_{u})\right]h_{u}(\overline{V}_{u})\lambda_{T}(u\mid\overline{V}_{u})Y_{T}(u)\de u\right)\\
=\E\left(\int\E\left(\left[h_{u}\{\overline{V}_{u},H(\psi^{*})\}-h_{u}^{*}(\overline{V}_{u})\right]Y_{T}(u)\mid\overline{V}_{u}\right)h_{u}(\overline{V}_{u})\lambda_{T}(u\mid\overline{V}_{u})\de u\right)=0,
\end{multline*}
for any $h_{u}(\overline{V}_{u})$. Because $h_{u}(\overline{V}_{u})$
is arbitrary, we obtain
\begin{equation}
\E\left(\left[h_{u}\{\overline{V}_{u},H(\psi^{*})\}-h_{u}^{*}(\overline{V}_{u})\right]Y_{T}(u)\mid\overline{V}_{u}\right)=0.\label{eq:eq2}
\end{equation}
Solving (\ref{eq:eq2}) for $h_{u}^{*}(\overline{V}_{u})$, we obtain
\[
h_{u}^{*}(\overline{V}_{u})=\E\left[h_{u}\{\overline{V}_{u},H(\psi^{*})\}\mid\overline{V}_{u},T\geq u\right].
\]
This completes the proof.

\subsection{Proof of Theorem \ref{Thm:projection}}

For any $B=B(F)$, let
\begin{multline*}
G=G(F)=\int_{0}^{\tau}\left[\E\left\{ B\dot{H}_{u}(\psi^{*})\mid\overline{V}_{u},T=u\right\} -\E\left\{ B\dot{H}_{u}(\psi^{*})\mid\overline{V}_{u},T\geq u\right\} \right]\\
\times\left[\var\left\{ H(\psi^{*})\mid\overline{V}_{u},T\geq u\right\} \right]^{-1}\left[H(\psi^{*})-\E\left\{ H(\psi^{*})\mid\overline{V}_{u},T\geq u\right\} \right]\de M_{T}(u).
\end{multline*}
To show $\prod\left(B\mid\Lambda^{\bot}\right)=G$, it is easy to
see that $G\in\Lambda^{\bot},$ so the remaining is to show that $B-G\in\Lambda$.
Toward this end, we show that for any $\tilde{G}=\tilde{G}(F)=\int_{0}^{\tau}\tilde{c}(\overline{V}_{u})[H(\psi^{*})-\E\{H(\psi^{*})\mid\overline{V}_{u},T\geq u\}]Y_{T}(u)\de M_{T}(u)\in\Lambda^{\bot}$,
$(B-G)\indep\tilde{G}$ or $\E\{(B-G)\tilde{G}\}=0$. We now verify
$\E(B\tilde{G})=\E(G\tilde{G})$ by the following calculation.

First, by (\ref{eq:lemma1}), we calculate 
\begin{eqnarray}
\E\left(G\tilde{G}\right) & = & \E\int_{0}^{\tau}\tilde{c}(\overline{V}_{u})[\E\{B\dot{H}_{u}(\psi^{*})\mid\overline{V}_{u},T=u\}-\E\{B\dot{H}_{u}(\psi^{*})\mid\overline{V}_{u},T\geq u\}]\nonumber \\
 &  & \times[\var\{H(\psi^{*})\mid\overline{V}_{u},T\geq u\}]^{-1}[H(\psi^{*})-\E\{H(\psi^{*})\mid\overline{V}_{u},T\geq u\}]{}^{2}\nonumber \\
 &  & \times\lambda_{T}(u\mid\overline{V}_{u})Y_{T}(u)\de u\nonumber \\
 & = & \E\int_{0}^{\tau}\tilde{c}(\overline{V}_{u})[\E\{B\dot{H}_{u}(\psi^{*})\mid\overline{V}_{u},T=u\}-\E\{B\dot{H}_{u}(\psi^{*})\mid\overline{V}_{u},T\geq u\}]\nonumber \\
 &  & \times\lambda_{T}(u\mid\overline{V}_{u})Y_{T}(u)\de u.\label{eq:right}
\end{eqnarray}
Second, we calculate 
\begin{eqnarray}
\E\left(B\tilde{G}\right) & = & \E\int_{0}^{\tau}\tilde{c}(\overline{V}_{u})B[H(\psi^{*})-\E\{H(\psi^{*})\mid\overline{L}_{u},T\geq u\}]\de M_{T}(u)\nonumber \\
 & = & \E\int_{0}^{\tau}\tilde{c}(\overline{V}_{u})B\dot{H}_{u}(\psi^{*})\de N_{T}(u)\nonumber \\
 &  & -\E\int_{0}^{\tau}\tilde{c}(\overline{V}_{u})B\dot{H}_{u}(\psi^{*})\lambda_{T}(u\mid\overline{V}_{u})Y_{T}(u)\de u\nonumber \\
 & = & \E\int_{0}^{\tau}\tilde{c}(\overline{V}_{u})[\E\{B\dot{H}_{u}(\psi^{*})\mid\overline{V}_{u},T=u\}-\E\{B\dot{H}_{u}(\psi^{*})\mid\overline{V}_{u},T\geq u\}]\nonumber \\
 &  & \times\lambda_{T}(u\mid\overline{V}_{u})Y_{T}(u)\de u,\label{eq:left}
\end{eqnarray}
where the last equality follows because 
\begin{multline*}
\E\int_{0}^{\tau}\tilde{c}(\overline{V}_{u})B\dot{H}_{u}(\psi^{*})\de N_{T}(u)=\E\int_{0}^{\tau}\tilde{c}(\overline{V}_{u})\E\{B\dot{H}_{u}(\psi^{*})\mid\overline{V}_{u},T\}\de N_{T}(u)\\
=\E\int_{0}^{\tau}\tilde{c}(\overline{V}_{u})\E\{B\dot{H}_{u}(\psi^{*})\mid\overline{V}_{u},T=u\}\lambda_{T}(u\mid\overline{V}_{u})Y_{T}(u)\de u,
\end{multline*}
and 
\begin{eqnarray*}
 &  & \E\left\{ \int_{0}^{\tau}\tilde{c}(\overline{V}_{u})B\dot{H}_{u}(\psi^{*})\lambda_{T}(u\mid\overline{V}_{u})Y_{T}(u)\de u\right\} \\
 & = & \E\left[\int_{0}^{\tau}\tilde{c}(\overline{V}_{u})\E\{B\dot{H}_{u}(\psi^{*})\mid\overline{V}_{u},Y_{T}(u)\}\lambda_{T}(u\mid\overline{V}_{u})Y_{T}(u)\de u\right]\\
 & = & \E\left[\int_{0}^{\tau}\tilde{c}(\overline{V}_{u})\E\{B\dot{H}_{u}(\psi^{*})\mid\overline{V}_{u},T\geq u\}\lambda_{T}(u\mid\overline{V}_{u})Y_{T}(u)\de u\right].
\end{eqnarray*}
Therefore, by (\ref{eq:right}) and (\ref{eq:left}), $\E(B\tilde{G})=\E(G\tilde{G})$
for any $\tilde{G}\in\Lambda^{\bot}$, proving (\ref{eq:projection}).

\subsection{Proof of Theorem \ref{Thm: efficient score}}

The semiparametric efficient score is $S_{\mathrm{eff}}^{*}(\psi^{*})=\prod\left(S_{\psi}\mid\Lambda^{\bot}\right)$.
By Theorem \ref{Thm:projection}, we have 
\begin{eqnarray*}
S_{\mathrm{eff}}^{*}(\psi^{*}) & = & \int_{0}^{\tau}[\E\{S_{\psi}\dot{H}_{u}(\psi^{*})\mid\overline{V}_{u},T=u\}-\E\{S_{\psi}\dot{H}_{u}(\psi^{*})\mid\overline{V}_{u},T\geq u\}]\\
 &  & \times[\var\{H(\psi^{*})\mid\overline{V}_{u},T\geq u\}]{}^{-1}[H(\psi^{*})-\E\{H(\psi^{*})\mid\overline{V}_{u},T\geq u\}]\de M_{T}(u)\\
 & = & -\int_{0}^{\tau}[\E\{\partial\dot{H}_{u}(\psi^{*})/\partial\psi\mid\overline{V}_{u},T=u\}-\E\{\partial\dot{H}_{u}(\psi^{*})/\partial\psi\mid\overline{V}_{u},T\geq u\}]\\
 &  & \times[\var\{H(\psi^{*})\mid\overline{V}_{u},T\geq u\}]{}^{-1}[H(\psi^{*})-\E\{H(\psi^{*})\mid\overline{V}_{u},T\geq u\}]\de M_{T}(u)\\
 & = & -\int_{0}^{\tau}\E\{\partial\dot{H}_{u}(\psi^{*})/\partial\psi\mid\overline{V}_{u},T=u\}[\var\{H(\psi^{*})\mid\overline{V}_{u},T\geq u\}]^{-1}\\
 &  & \times H(\psi^{*})-\E\{H(\psi^{*})\mid\overline{V}_{u},T\geq u\}]\de M_{T}(u),
\end{eqnarray*}
where the last equality follows by using the generalized information
equality: because $\dot{H}_{u}(\psi^{*})=H(\psi^{*})-\E\{H(\psi^{*})\mid\overline{V}_{u},T\geq u\}$,
we have $\E\{\dot{H}_{u}(\psi^{*})\mid\overline{V}_{u},T\geq u\}=0$.
Take the derivative of $\psi$ at both sides, we have $\E\{S_{\psi}\dot{H}_{u}(\psi^{*})\mid\overline{V}_{u},T\geq u\}+\E\{\partial\dot{H}_{u}(\psi^{*})/\partial\psi\mid\overline{V}_{u},T\geq u\}=0$,
or equivalently $\E\{S_{\psi}\dot{H}_{u}(\psi^{*})\mid\overline{V}_{u},T\geq u\}=-\E\{\partial\dot{H}_{u}(\psi^{*})/\partial\psi\mid\overline{V}_{u},T\geq u\}$.
Similarly, noticing $\E\{H(\psi^{*})\mid\overline{V}_{u},T\geq u\}=\E\{H(\psi^{*})\mid\overline{V}_{u},T=u\}$,
we have $\E\{\dot{H}_{u}(\psi^{*})\mid\overline{V}_{u},T=u\}=0$.
Take the derivative of $\psi$ at both sides, we have $\E\{S_{\psi}\dot{H}_{u}(\psi^{*})\mid\overline{V}_{u},T=u\}+\E\{\partial\dot{H}_{u}(\psi^{*})/\partial\psi\mid\overline{V}_{u},T=u\}=0$,
or equivalently $\E\{S_{\psi}\dot{H}_{u}(\psi^{*})\mid\overline{V}_{u},T=u\}=-\E\{\partial\dot{H}_{u}(\psi^{*})/\partial\psi\mid\overline{V}_{u},T=u\}$.
Ignoring the negative sign, the result in Theorem \ref{Thm: efficient score}
follow.

\subsection{Proof of Theorem \ref{Thm:2-dr} \label{subsec:Proof-of-dr}}

We show that $\E\{G(\psi^{*};F,c)\}=0$ in two cases.

First, if $\lambda_{T}(t\mid\overline{V}_{t})$ is correctly specified,
under Assumption \ref{asumption:CT-UNC}, $M_{T}(t)$ is a martingale
with respect to the filtration $\sigma\{\overline{V}_{t},H(\psi^{*})\}$.
Because $c(\overline{V}_{u})[H(\psi^{*})-\E\{H(\psi^{*})\mid\overline{V}_{u},T\geq u\}]$
is a $\sigma\{\overline{V}_{t},H(\psi^{*})\}$-predictable process,
$\int_{0}^{t}c(\overline{V}_{u})[H(\psi^{*})-\E\{H(\psi^{*})\mid\overline{V}_{u},T\geq u\}]\de M_{T}(u)$
is a martingale for $t\geq0$. Therefore, $\E\{G(\psi^{*};F,c)\}=0$.

Second, if $\E\left\{ H(\psi^{*})\mid\overline{V}_{u},T\geq u\right\} $
is correctly specified but $\lambda_{T}(t\mid\overline{V}_{t})$ is
not necessarily correctly specified, let $\lambda_{T}^{*}(t\mid\overline{V}_{t})$
be the probability limit of the possibly misspecified model. We obtain
\begin{eqnarray}
 &  & \E\int c(\overline{V}_{u})\left[H(\psi^{*})-\E\left\{ H(\psi^{*})\mid\overline{V}_{u},T\geq u;\beta^{*}\right\} \right]\left\{ \de N_{T}(u)-\lambda_{T}^{*}(u\mid\overline{V}_{u})Y_{T}(u)\de u\right\} \nonumber \\
 & = & \E\int c(\overline{V}_{u})\left[H(\psi^{*})-\E\left\{ H(\psi^{*})\mid\overline{V}_{u},T\geq u;\beta^{*}\right\} \right]\left\{ \de N_{T}(u)-\lambda_{T}(u\mid\overline{V}_{u})Y_{T}(u)\de u\right\} \nonumber \\
 &  & +\E\int c(\overline{V}_{u})\left[H(\psi^{*})-\E\left\{ H(\psi^{*})\mid\overline{V}_{u},T\geq u;\beta^{*}\right\} \right]\left\{ \lambda_{T}(u\mid\overline{V}_{u})-\lambda_{T}^{*}(u\mid\overline{V}_{u})\right\} Y_{T}(u)\de u\nonumber \\
 & = & 0+\E\int c(\overline{V}_{u})E\left(\left[H(\psi^{*})-\E\left\{ H(\psi^{*})\mid\overline{V}_{u},T\geq u;\beta^{*}\right\} \right]\mid\overline{V}_{u},T\geq u\right)\label{eq:eq3}\\
 &  & \times\left\{ \lambda_{T}(u\mid\overline{V}_{u})-\lambda_{T}^{*}(u\mid\overline{V}_{u})\right\} Y_{T}(u)\de u\nonumber \\
 & = & 0+\E\int c(\overline{V}_{u})\times0\times\left\{ \lambda_{T}(u\mid\overline{V}_{u})-\lambda_{T}^{*}(u\mid\overline{V}_{u})\right\} Y_{T}(u)\de u=0,\label{eq:eq4}
\end{eqnarray}
where zero in (\ref{eq:eq3}) follows because $\de M_{T}(u)=\de N_{T}(u)-\lambda_{T}(u\mid\overline{V}_{u})\de u$
is a martingale with respect to the filtration $\sigma\{\overline{V}_{t},H(\psi^{*})\}$,
and zero in (\ref{eq:eq4}) follows because $\E\left\{ H(\psi^{*})\mid\overline{V}_{u},T\geq u\right\} $
is correctly specified and therefore, $\E\left\{ H(\psi^{*})\mid\overline{V}_{u},T\geq u;\beta^{*}\right\} =\E\left\{ H(\psi^{*})\mid\overline{V}_{u},T\geq u\right\} $.

\subsection{Proof of Theorem \ref{Thm:3-mr}}

We show that $\E\left\{ \delta_{C}G(\psi^{*},\beta^{*},M_{T}^{*};F)/K_{C}^{*}(\tau\mid\overline{V}_{\tau})\right\} =0$
in three cases.

First, under Scenarios (a), (b), and (c) listed in Table \ref{tab:Multiply-Robustness}
when $K_{C}^{*}$ is correctly specified for $K_{C}$, either $\E\{H(\psi^{*})\mid\overline{V}_{u},T\geq u;\beta^{*}\}$
is correctly specified or $M_{T}^{*}$ is correctly specified for
$M_{T}$, we show that (\ref{eq:IPCW ee}) is an unbiased estimating
equation. Under these scenarios, we have $K_{C}^{*}(\tau\mid\overline{V}_{\tau})=K_{C}(\tau\mid\overline{V}_{\tau})$.
It suffices to show that 
\begin{eqnarray*}
\E\left\{ \frac{\delta_{C}}{K_{C}\left(\tau\mid\overline{V}_{\tau}\right)}G(\psi^{*},\beta^{*},M_{T}^{*};F)\right\}  & = & \E\left[\E\left\{ \frac{\delta_{C}}{K_{C}\left(\tau\mid\overline{V}_{\tau}\right)}G(\psi^{*},\beta^{*},M_{T}^{*};F)\mid F\right\} \right]\\
 & = & \E\left\{ \frac{\E(\delta_{C}\mid F)}{K_{C}\left(\tau\mid\overline{V}_{\tau}\right)}G(\psi^{*},\beta^{*},M_{T}^{*};F)\right\} \\
 & = & \E\left[\frac{\E(\delta_{C}\mid\overline{V}_{\tau})}{K_{C}\left(\tau\mid\overline{V}_{\tau}\right)}\E\{G(\psi^{*},\beta^{*},M_{T}^{*};F)\mid\overline{V}_{\tau}\}\right],
\end{eqnarray*}
where the third equality follows by Assumption \ref{asp:NUC-1}, and
the last equality follows by Theorem \ref{Thm:2-dr}.

Second, under Scenarios (b) and (d) listed in Table \ref{tab:Multiply-Robustness}
when $\E\{H(\psi^{*})\mid\overline{V}_{u},T\geq u;\beta^{*}\}$ is
correctly specified, we have $\E\{H(\psi^{*})\mid\overline{V}_{u},T\geq u;\beta^{*}\}=\E\{H(\psi^{*})\mid\overline{V}_{u},T\geq u\}$.
Also, under Assumption \ref{asumption:CT-UNC}, $\E\left\{ H(\psi^{*})\mid\overline{V}_{u},T\geq u\right\} =\E\left\{ H(\psi^{*})\mid\overline{V}_{u}\right\} $.
Then, we have
\begin{eqnarray*}
 &  & \E\{G(\psi^{*},\beta^{*},M_{T}^{*};F)\mid\overline{V}_{\tau}\}\\
 & = & \E\left\{ \int c(\overline{V}_{u})\left[H(\psi^{*})-\E\left\{ H(\psi^{*})\mid\overline{V}_{u},T\geq u;\beta^{*}\right\} \right]\de M_{T}^{*}(u)\mid\overline{V}_{\tau}\right\} \\
 & = & \int c(\overline{V}_{u})\E\left[H(\psi^{*})-\E\left\{ H(\psi^{*})\mid\overline{V}_{u},T\geq u\right\} \mid\overline{V}_{\tau}\right]\E\left\{ \de M_{T}^{*}(u)\mid\overline{V}_{\tau}\right\} \\
 & = & \int c(\overline{V}_{u})\E\left[\E\left\{ H(\psi^{*})\mid\overline{V}_{u}\right\} -\E\left\{ H(\psi^{*})\mid\overline{V}_{u},T\geq u\right\} \mid\overline{V}_{\tau}\right]\E\left\{ \de M_{T}^{*}(u)\mid\overline{V}_{\tau}\right\} \\
 & = & \int c(\overline{V}_{u})\E\left[\E\left\{ H(\psi^{*})\mid\overline{V}_{u},T\geq u\right\} -\E\left\{ H(\psi^{*})\mid\overline{V}_{u},T\geq u\right\} \mid\overline{V}_{\tau}\right]\E\left\{ \de M_{T}^{*}(u)\mid\overline{V}_{\tau}\right\} =0.
\end{eqnarray*}
It follows that
\begin{eqnarray*}
\E\left\{ \frac{\delta_{C}}{K_{C}^{*}\left(\tau\mid\overline{V}_{\tau}\right)}G(\psi^{*},\beta^{*},M_{T}^{*};F)\right\}  & = & \E\left[\E\left\{ \frac{\delta_{C}}{K_{C}^{*}\left(\tau\mid\overline{V}_{\tau}\right)}G(\psi^{*},\beta^{*},M_{T}^{*};F)\mid F\right\} \right]\\
 & = & \E\left[\frac{K_{C}\left(\tau\mid\overline{V}_{\tau}\right)}{K_{C}^{*}\left(\tau\mid\overline{V}_{\tau}\right)}\E\{G(\psi^{*},\beta^{*},M_{T}^{*};F)\mid\overline{V}_{\tau}\}\right]\\
 & = & \E\left\{ \frac{K_{C}\left(\tau\mid\overline{V}_{\tau}\right)}{K_{C}^{*}\left(\tau\mid\overline{V}_{\tau}\right)}\times0\right\} =0.
\end{eqnarray*}

Third, under Scenario (e) listed in Table \ref{tab:Multiply-Robustness}
when $M_{T}^{*}$ is correctly specified for $M_{T}$, we have
\begin{equation}
\E\left\{ \de M_{T}(u)\mid\overline{V}_{u}\right\} =0,\ (u>0).\label{eq:dM}
\end{equation}
Define $\kappa(\overline{V}_{u})=E\left\{ K_{C}\left(\tau\mid\overline{V}_{\tau}\right)/K_{C}^{*}\left(\tau\mid\overline{V}_{\tau}\right)\mid\overline{V}_{u}\right\} $
for all $u>0$. We show $\E\left\{ \delta_{C}G(\psi^{*},\beta^{*},M_{T};F)/K_{C}^{*}(\tau\mid\overline{V}_{\tau})\right\} =0$
by induction. Let $\Delta>0$ be a small increment. We start with
\begin{eqnarray}
 &  & \E\left\{ \frac{\delta_{C}}{K_{C}^{*}\left(\tau\mid\overline{V}_{\tau}\right)}G(\psi^{*},\beta^{*},M_{T};F)\right\} \nonumber \\
 & = & \E\left[\kappa(\overline{V}_{\tau})\int_{0}^{\tau}c(\overline{V}_{u})\left[H(\psi^{*})-\E\left\{ H(\psi^{*})\mid\overline{V}_{u},T\geq u;\beta^{*}\right\} \right]\de M_{T}(u)\right]\label{eq:(tau)}\\
 & = & \E\left[\kappa(\overline{V}_{\tau})\E\left\{ \left(\int_{0}^{\tau-\Delta}+\int_{\tau-\Delta}^{\tau}\right)c(\overline{V}_{u})\left[H(\psi^{*})-\E\left\{ H(\psi^{*})\mid\overline{V}_{u},T\geq u;\beta^{*}\right\} \right]\de M_{T}(u)\mid\overline{V}_{\tau}\right\} \right]\nonumber \\
 & = & \E\left[\E\left\{ \kappa(\overline{V}_{\tau})\mid\overline{V}_{\tau-\Delta}\right\} \int_{0}^{\tau-\Delta}c(\overline{V}_{u})\left[\E\left\{ H(\psi^{*})\mid\overline{V}_{u},T\geq u\right\} -\E\left\{ H(\psi^{*})\mid\overline{V}_{u},T\geq u;\beta^{*}\right\} \right]\de M_{T}(u)\right]\nonumber \\
 &  & +\E\left\{ \kappa(\overline{V}_{\tau})\int_{\tau-\Delta}^{\tau}c(\overline{V}_{u})\left[\E\left\{ H(\psi^{*})\mid\overline{V}_{u},T\geq u\right\} -\E\left\{ H(\psi^{*})\mid\overline{V}_{u},T\geq u;\beta^{*}\right\} \right]\de M_{T}(u)\right\} \nonumber \\
 & = & \E\left\{ \kappa(\overline{V}_{\tau-\Delta})\int_{0}^{\tau-\Delta}c(\overline{V}_{u})\left[\E\left\{ H(\psi^{*})\mid\overline{V}_{u},T\geq u\right\} -\E\left\{ H(\psi^{*})\mid\overline{V}_{u},T\geq u;\beta^{*}\right\} \right]\de M_{T}(u)\right\} \nonumber \\
 &  & +\E\left\{ \kappa(\overline{V}_{\tau})c(\overline{V}_{\tau})\left[\E\left\{ H(\psi^{*})\mid\overline{V}_{\tau},T\geq\tau\right\} -\E\left\{ H(\psi^{*})\mid\overline{V}_{\tau},T\geq\tau;\beta^{*}\right\} \right]\E\left\{ \de M_{T}(\tau)\mid\overline{V}_{\tau}\right\} \right\} \nonumber \\
 & = & \E\left\{ \left[\kappa(\overline{V}_{\tau-\Delta})\int_{0}^{\tau-\Delta}c(\overline{V}_{u})\left[\E\left\{ H(\psi^{*})\mid\overline{V}_{u},T\geq u\right\} -\E\left\{ H(\psi^{*})\mid\overline{V}_{u},T\geq u;\beta^{*}\right\} \right]\de M_{T}(u)\right]\right\} +0,\label{eq:(tau-)}
\end{eqnarray}
where $0$ in the last equality follows by (\ref{eq:dM}). Note that
(\ref{eq:(tau-)}) is (\ref{eq:(tau)}) replacing $\tau$ by $\tau-\Delta$.
We then repeat the same calculation for (\ref{eq:(tau-)}) until $\tau-\Delta$
reaches zero. The last step is to recognize that (\ref{eq:(tau-)})
with $\tau-\Delta=0$ is zero. This completes the proof.

\subsection{Proof of Theorem \ref{thm:4}}

We assume the multiple robustness condition holds; i.e. either the
potential outcome mean model or the model for the treatment process
is correctly specified, regardless the model for the censoring process
is correctly specified.

Taylor expansion of $\bP_{n}\left\{ \Phi(\widehat{\psi},\widehat{\beta},\widehat{M}_{T},\widehat{K}_{C};F)\right\} =0$
around $\psi^{*}$ leads to 
\begin{eqnarray*}
0 & = & \bP_{n}\left\{ \Phi(\widehat{\psi},\widehat{\beta},\widehat{M}_{T},\widehat{K}_{C};F)\right\} \\
 & = & \bP_{n}\left\{ \Phi(\psi^{*},\widehat{\beta},\widehat{M}_{T},\widehat{K}_{C};F)\right\} +\bP_{n}\left\{ \frac{\partial\Phi(\widetilde{\psi},\widehat{\beta},\widehat{M}_{T},\widehat{K}_{C};F)}{\partial\psi^{\T}}\right\} (\widehat{\psi}-\psi^{*}),
\end{eqnarray*}
where $\widetilde{\psi}$ is on the line segment between $\widehat{\psi}$
and $\psi^{*}$.

Under Assumption \ref{asump:donsker} (i) and (ii), 
\[
(\bP_{n}-\bP)\left\{ \frac{\partial\Phi(\widetilde{\psi},\widehat{\beta},\widehat{M}_{T},\widehat{K}_{C};F)}{\partial\psi^{\T}}\right\} =(\bP_{n}-\bP)\left\{ \frac{\partial\Phi(\psi^{*},\beta^{*},M_{T}^{*},K_{C}^{*};F)}{\partial\psi^{\T}}\right\} =o_{p}(n^{-1/2}),
\]
and therefore, 
\begin{eqnarray*}
\bP_{n}\left\{ \frac{\partial\Phi(\widetilde{\psi},\widehat{\beta},\widehat{M}_{T},\widehat{K}_{C};F)}{\partial\psi^{\T}}\right\}  & = & \bP\left\{ \frac{\partial\Phi(\widetilde{\psi},\widehat{\beta},\widehat{M}_{T},\widehat{K}_{C};F)}{\partial\psi^{\T}}\right\} +o_{p}(n^{-1/2})\\
 & = & A(\psi^{*},\beta^{*},M_{T}^{*},K_{C}^{*})+o_{p}(n^{-1/2}).
\end{eqnarray*}
We then have 
\begin{equation}
n^{1/2}(\widehat{\psi}-\psi^{*})=\left\{ A(\psi^{*},\beta^{*},M_{T}^{*},K_{C}^{*})\right\} ^{-1}n^{1/2}\bP_{n}\left\{ \Phi(\psi^{*},\widehat{\beta},\widehat{M}_{T},\widehat{K}_{C};F)\right\} +o_{p}(1).\label{eq:1}
\end{equation}
Based on the multiple robustness, we have
\begin{equation}
\bP\{\Phi(\psi^{*},\beta^{*},M_{T}^{*},K_{C}^{*};F)\}=0.\label{eq:2-1}
\end{equation}
To express (\ref{eq:1}) further, based on (\ref{eq:2-1}), we have
\begin{multline}
\bP_{n}\Phi(\psi^{*},\widehat{\beta},\widehat{M}_{T},\widehat{K}_{C};F)=(\bP_{n}-\bP)\Phi(\psi^{*},\widehat{\beta},\widehat{M}_{T},\widehat{K}_{C};F)\\
+\bP\left\{ \Phi(\psi^{*},\widehat{\beta},\widehat{M}_{T},\widehat{K}_{C};F)-\Phi(\psi^{*},\beta^{*},M_{T}^{*},K_{C}^{*};F)\right\} +\bP\Phi(\psi^{*},\beta^{*},M_{T}^{*},K_{C}^{*};F).\label{eq:2}
\end{multline}
By Assumption \ref{asump:donsker} (i) and (ii), the first term in
(\ref{eq:eq2}) becomes 
\begin{eqnarray}
(\bP_{n}-\bP)\Phi(\psi^{*},\widehat{\beta},\widehat{M}_{T},\widehat{K}_{C};F) & = & (\bP_{n}-\bP)\Phi(\psi^{*},\beta^{*},M_{T}^{*},K_{C}^{*};F)+o_{p}(n^{-1/2})\nonumber \\
 & = & \bP_{n}\Phi(\psi^{*},\beta^{*},M_{T}^{*},K_{C}^{*};F)+o_{p}(n^{-1/2}).\label{eq:2-2}
\end{eqnarray}
By Assumption \ref{asump:donsker} (iv), the second term in (\ref{eq:eq2})
becomes 
\begin{eqnarray}
 &  & \bP\left\{ \Phi(\psi^{*},\widehat{\beta},\widehat{M}_{T},\widehat{K}_{C};F)-\Phi(\psi^{*},\beta^{*},M_{T}^{*},K_{C}^{*};F)\right\} \nonumber \\
 & = & J(\widehat{\beta},\widehat{M}_{T},\widehat{K}_{C})-J(\beta^{*},M_{T}^{*},K_{C}^{*})+o_{p}(n^{-1/2})\nonumber \\
 & = & J_{1}(\widehat{\beta})-J_{1}(\beta^{*})+J_{2}(\widehat{M}_{T})-J_{2}(M_{T}^{*})+J_{3}(\widehat{K}_{C})-J_{3}(M_{C}^{*})+o_{p}(n^{-1/2})\nonumber \\
 & = & \bP_{n}\Phi_{1}(\psi^{*},\beta^{*},M_{T}^{*},K_{C}^{*};F)+\bP_{n}\Phi_{2}(\psi^{*},\beta^{*},M_{T}^{*},K_{C}^{*};F)\nonumber \\
 &  & +\bP_{n}\Phi_{3}(\psi^{*},\beta^{*},M_{T}^{*},K_{C}^{*};F).\label{eq:2-3}
\end{eqnarray}
Combining (\ref{eq:2-1})\textendash (\ref{eq:2-3}), 
\[
\bP_{n}\Phi(\psi^{*},\widehat{\beta},\widehat{M}_{T},\widehat{K}_{C};F)=\bP_{n}\{\widetilde{B}(\psi^{*},\beta^{*},M_{T}^{*},K_{C}^{*};F)\},
\]
where 
\begin{eqnarray*}
\widetilde{B}(\psi^{*},\beta^{*},M_{T}^{*},K_{C}^{*};F) & = & \Phi(\psi^{*},\beta^{*},M_{T}^{*},K_{C}^{*};F)+\Phi_{1}(\psi^{*},\beta^{*},M_{T}^{*},K_{C}^{*};F)\\
 &  & +\Phi_{2}(\psi^{*},\beta^{*},M_{T}^{*},K_{C}^{*};F)+\Phi_{3}(\psi^{*},\beta^{*},M_{T}^{*},K_{C}^{*};F).
\end{eqnarray*}
As a result, 
\begin{equation}
n^{1/2}(\widehat{\psi}-\psi^{*})=n^{1/2}\bP_{n}\widetilde{\Phi}(\psi^{*},\beta^{*},K_{V}^{*},K_{C}^{*};F)+o_{p}(1),\label{eq:(2.3)}
\end{equation}
where
\[
\widetilde{\Phi}(\psi^{*},\beta^{*},M_{T}^{*},K_{C}^{*};F)=\left\{ A(\psi^{*},\beta^{*},M_{T}^{*},K_{C}^{*})\right\} ^{-1}\widetilde{B}(\psi^{*},\beta^{*},M_{T}^{*},K_{C}^{*};F).
\]

We now consider the case when all nuisance models are correctly specified,
i.e., $\E\{H(\psi^{*})\mid\overline{V}_{u},T\geq u;\beta^{*}\}=\E\{H(\psi^{*})\mid\overline{V}_{u},T\geq u\}$,
$M_{T}^{*}=M_{T},$ and $K_{C}^{*}=K_{C}$. 

Define the score functions: $S_{\beta}=S_{\beta}\{H(\psi^{*}),\overline{V}_{u},T\geq u\}$.
Then, the tangent space for $\beta$ is $\widetilde{\Lambda}_{1}=\{S_{\beta}\in\R^{p}:\E(S_{\beta}\mid\overline{V}_{u},T\geq u)=0\}$.
Following \citet{tsiatis2007semiparametric}, the nuisance tangent
space for the proportional hazards model (\ref{eq:ph-V}) is 
\[
\widetilde{\Lambda}_{2}=\left\{ S_{\alpha}+\int h(u)\de M_{T}(u):\ h(u)\in\mathbb{\R}^{p}\right\} ,
\]
where
\begin{equation}
S_{\alpha}\coloneqq\int\left\{ W_{T}(u,\overline{V}_{u})-\frac{\E\left[W_{T}(u,\overline{V}_{u})\exp\left\{ \gamma^{\T}W_{T}(u,\overline{V}_{u})\right\} Y_{T}(u)\right]}{\E\left[\exp\left\{ \alpha^{\T}W_{T}(u,\overline{V}_{u})\right\} Y_{T}(u)\right]}\right\} \de M_{T}(u).\label{eq:S_alpha}
\end{equation}
The nuisance tangent space for the proportional hazards model (\ref{eq:ph-C})
is 
\[
\widetilde{\Lambda}_{3}=\left\{ S_{\eta}+\int h(u)\de M_{C}(u):\ h(u)\in\mathbb{\R}^{p}\right\} ,
\]
where
\begin{equation}
S_{\eta}\coloneqq\int\left\{ W_{C}(u,\overline{V}_{u})-\frac{\E\left[W_{C}(u,\overline{V}_{u})\exp\left\{ \eta^{\T}W_{C}(u,\overline{V}_{u})\right\} Y_{C}(u)\right]}{\E\left[\exp\left\{ \eta^{\T}W_{C}(u,\overline{V}_{u})\right\} Y_{C}(u)\right]}\right\} \de M_{C}(u).\label{eq:S_eta}
\end{equation}
Assuming that the treatment process and the censoring process can
not jump at the same time point, $\widetilde{\Lambda}_{1}$, $\widetilde{\Lambda}_{2}$,
and $\widetilde{\Lambda}_{3}$ are mutually orthogonal to each other.
Therefore, the nuisance tangent space for $\beta$ and the proportional
hazards models (\ref{eq:ph-V}) and (\ref{eq:ph-C}) is $\widetilde{\Lambda}=\widetilde{\Lambda}_{1}\oplus\widetilde{\Lambda}_{2}\oplus\widetilde{\Lambda}_{3}$.
The influence function for $\widehat{\psi}$ is
\begin{eqnarray}
 &  & \widetilde{B}(\psi^{*},\beta^{*},M_{T},K_{C};F)\nonumber \\
 & = & \Phi(\psi^{*},\beta^{*},M_{T},K_{C};F)-\prod\left\{ \Phi(\psi^{*},\beta^{*},M_{T},K_{C};F)\mid\widetilde{\Lambda}\right\} \nonumber \\
 & = & \Phi(\psi^{*},\beta^{*},M_{T},K_{C};F)-E\left\{ \Phi(\psi^{*},\beta^{*},M_{T},K_{C};F)S_{\alpha}^{\T}\right\} \E\left(S_{\alpha}S_{\alpha}^{\T}\right)^{-1}S_{\alpha}\nonumber \\
 &  & -\E\left\{ \Phi(\psi^{*},\beta^{*},M_{T},K_{C};F)S_{\eta}^{\T}\right\} \E\left(S_{\eta}S_{\eta}^{\T}\right)^{-1}S_{\eta}\nonumber \\
 &  & +\int\frac{\E\left[G(\psi^{*},\beta^{*},M_{T};F)\exp\left\{ \alpha^{\T}W_{T}(u,\overline{V}_{u})\right\} \delta_{C}/K_{C}(\tau\mid\overline{V}_{\tau})\right]}{\E\left[\exp\left\{ \alpha^{\T}W_{T}(u,\overline{V}_{u})\right\} Y_{T}(u)\right]}\de M_{C}(u)\nonumber \\
 &  & +\int\frac{\E\left[G(\psi^{*},\beta^{*},M_{T};F)\exp\left\{ \eta^{\T}W_{C}(u,\overline{V}_{u})\right\} \delta_{C}/K_{C}(\tau\mid\overline{V}_{\tau})\right]}{\E\left[\exp\left\{ \eta^{\T}W_{C}(u,\overline{V}_{u})\right\} Y_{C}(u)\right]}\de M_{T}(u).\label{eq:tilde-J}
\end{eqnarray}

\section{Details for the simulation study}

First, Algorithm \ref{tab:Algorithm-for-generating} specifies the
steps for generating $T$ according to a time-dependent proportional
hazards model. 
\begin{algorithm}
\caption{\label{tab:Algorithm-for-generating}Algorithm 1 for generating $T$
according to a time-dependent proportional hazards model}

\begin{description}
\item [{Step$\ 1.$}] Set $k=1$.
\item [{Step$\ 2.$}] Generate a temporary time to treatment initiation,
$T_{\temp,k}$, compatible with the hazard function for the time interval
$[t_{k},t_{k+1})$, using the method of \citet{bender2005generating};
i.e., generate $u\sim$Uniform$[0,1]$ and let $T_{\temp,k}=$$-\log(1-u)/\{\lambda_{T,0}\exp(\alpha_{1}^{*}L_{TI}+\alpha_{2}^{*}L_{TD,t_{k}})\}$.
\begin{description}
\item [{If}] $T_{\temp,k}$ is contained within the first time interval
$[0,t_{k+1}-t_{k})$, then set $T=T_{\temp,k}+t_{k};$
\item [{else$\ $if}] $T_{\temp,k}$ is not contained within the interval
$[0,t_{k+1}-t_{k})$, increase $k$ by $1$ and move to the beginning
of Step 2.
\end{description}
\end{description}
\end{algorithm}
 Second, we describe the nuisance models and their estimation. For
$c(\overline{V}_{u})$, we approximate $\E\{(1,T)^{\T}(\tau-T)I(T\leq\tau)\mid\overline{V}_{u},T\geq u\}$
by $\widehat{\pr}(T\leq\tau\mid\overline{V}_{u},T\geq u)\times\widehat{\E}\{(1,T)^{\T}(\tau-T)\mid\overline{V}_{u},u\leq T\leq\tau\}$.
We describe the details for fitting below:
\begin{description}
\item [{(a)}] $\widehat{\pr}(T\leq\tau\mid\overline{V}_{u},T\geq u)$ is
the predicted value from a logistic regression model of $I(T\leq\tau)$
against $u$, $L_{TI}$, $L_{TD,u}$, and all interactions of these
terms, restricted to subjects with $T\geq u$.
\item [{(b)}] $\widehat{\E}(\tau-T\mid\overline{V}_{u},u\leq T\leq\tau)$
is the predicted value from a linear regression model of $\tau-T$
against $u$, $L_{TI}$, $L_{TD,u}$, and all interactions of these
terms, restricted to subjects with $u\leq T\leq\tau$.
\item [{(c)$\widehat{\E}\{T(\tau-T)\mid\overline{V}_{u},u\leq T\leq\tau\}$}] is
the predicted value from a linear regression model of $T(\tau-T)$
against $u$, $L_{TI}$, $L_{TD,u}$, and all interactions of these
terms, restricted to subjects with $u\leq T\leq\tau$.
\item [{(d)}] $\E\{H(\widehat{\psi}_{p})\mid\overline{V}_{u},T\geq u;\widehat{\beta}\}$
by a linear regression model of $H(\widehat{\psi}_{p})$ against $u$,
$L_{TI}$, $L_{TD,u}$, and all interactions of these terms, restricted
to subjects with $T\geq u$.
\end{description}

\end{document}